\documentclass[pre,tighten]{revtex4}

\usepackage{amssymb,amsmath}

\usepackage{graphicx}

\usepackage{color}

\newcommand{\vicente}[1]{{ #1}}
\newcommand\beq{\begin{equation}}
\newcommand\eeq{\end{equation}}
\newcommand\beqa{\begin{eqnarray}}
\newcommand\eeqa{\end{eqnarray}}
\newcommand{\dd}{\text{d}}

\newcommand{\al}{\alpha}

\begin{document}

\title{Homogeneous steady states in a granular fluid driven by a stochastic bath with friction}
\author{Mois\'es G. Chamorro}
\author{Francisco Vega Reyes}
\email{fvega@unex.es} \homepage{http://www.unex.es/eweb/fisteor/fran/}
\affiliation{Departamento de F\'{\i}sica, Universidad de Extremadura, E-06071 Badajoz, Spain}
\author{Vicente Garz\'o}
\email{vicenteg@unex.es} \homepage{http://www.eweb.unex.es/eweb/fisteor/vicente/}
\affiliation{Departamento de F\'{\i}sica, Universidad de Extremadura, E-06071 Badajoz, Spain}

\begin{abstract}

The homogeneous state of a granular flow of smooth inelastic hard spheres or disks described by the Enskog-Boltzmann kinetic equation is analyzed. The granular gas is fluidized by the presence of a random force and a drag force. The combined action of both forces, that act homogeneously on the granular gas, tries to mimic the interaction of the set of particles with a surrounding fluid. The first stochastic force thermalizes the system, providing for the necessary energy recovery to keep the system in its gas state at all times, whereas the second force allows us to mimic the action of the surrounding fluid viscosity. After a transient regime, the gas reaches a steady state characterized by a \emph{scaled} distribution function $\varphi$ that does not only depend on the dimensionless velocity $\mathbf{c}\equiv \mathbf{v}/v_0$ ($v_0$ being the thermal velocity) but also on the dimensionless driving force parameters. The dependence of $\varphi$ and its first relevant velocity moments $a_2$ and $a_3$ (which measure non-Gaussian properties of $\varphi$) on both the coefficient of restitution $\al$ and the driven parameters is widely investigated by means of the direct simulation Monte Carlo method. In addition, approximate forms for $a_2$ and $a_3$ are also derived from an expansion of $\varphi$ in Sonine polynomials. The theoretical expressions of the above Sonine coefficients agree  well with simulation data, even for quite small values of $\alpha$.  Moreover, the third order expansion of the distribution function makes a significant accuracy improvement for larger velocities and inelasticities. Results also show that the non-Gaussian corrections to the distribution function $\varphi$ are smaller than those observed for undriven granular gases.
\end{abstract}


\date{\today}
\maketitle

\section{Introduction}
\label{sec1}

A large amount of work has been devoted in the last few decades to the study of granular matter (materials composed by many mesoscopic particles that collide inelastically). When the granular material is externally excited (rapid flow conditions), the behavior of solid particles is dominated by the particle collisions and kinetic theory tools can be used to describe granular flows. Thus, from the point of view of fundamental kinetic theory, the study of granular gases is interesting because it involves the generalization of classical kinetic equations (such as the Boltzmann, Enskog or  Boltzmann-Lorentz equations, for instance) to dissipative dynamics. On the other hand, the fact that collisions are inelastic gives rise to a decreasing time evolution of the total kinetic energy and so, one has to inject energy into the system to keep it under rapid flow conditions. When the injected energy compensates for the collisional loss of energy, the system reaches a \emph{non-equilibrium} steady state. In this context, granular matter can be considered as a good prototype of a system that inherently is in a non-equilibrium state.

In real experiments, the granular gas is driven through the boundaries, for example, vibrating its walls \cite{YHCMW02} or alternatively by bulk driving, as in air-fluidized beds \cite{AD06,SGS05}. The same effect can be achieved by means of the action of an \emph{external} driving force that heats the system homogeneously. This way of supplying energy is quite usual in computer simulations \cite{Puglisi,Zippelius} and this type of external forces is usually called ``thermostat'' \cite{E90}. Although thermostats have been widely used in the past to study granular dynamics, their effects on the properties of the system are not yet completely understood \cite{DSBR86,GSB90,GS03}.

In this paper, we are interested in analyzing the homogeneous steady state of a driven granular fluid. Our thermostat is composed by two different terms: (i) a drag force proportional to the velocity of the particle and (ii) a stochastic force with the form of a Gaussian white noise where the particles are randomly kicked between collisions \cite{WM96}. Under these conditions, our kinetic equation has the structure of a Fokker-Planck equation \cite{VK92} plus the corresponding (inelastic) collisional operator of the Enskog-Boltzmann equation. The viscous drag force allows us to model the friction from a surrounding fluid over a moderately dense set of spheres \cite{GTSH12}. The stochastic force would model the energy transfer from the surrounding fluid molecules to the granular particles, due to molecular thermal motion (much in the same way as in a Brownian particle). Thus, our study has obvious applications to the dynamics of colloids and suspensions \cite{J00,K90,KH01,BGP11,GTSH12}. In particular, and since the volume forces act homogeneously in the granular gas, our system may show homogeneous steady states, if there is no additional energy input from the boundaries.

The same type of thermostats were used in previous works by other authors \cite{Puglisi}. In particular, Gradenigo \emph{et al.} \cite{GSVP11} carried out Langevin dynamics simulations for hard disks to measure the static and dynamic structure factors for shear and longitudinal modes. The corresponding best fit of their simulation results allows them to identify the kinematic and longitudinal viscosities and the thermal diffusivity. For the sake of simplicity, they neglect non-Gaussian corrections to the (homogeneous) distribution function and use the forms of the \emph{elastic} Enskog transport coefficients to compare with simulations. More recently, the expressions of the \emph{inelastic} transport coefficients of driven granular fluids have been derived \cite{GCV13} by means of the Chapman-Enskog method instead \cite{CC70}. In this case, the inherently homogeneous steady state of our system emerges as the zeroth-order approximation $f^{(0)}$ in the Chapman-Enskog perturbative scheme. In order to characterize the deviation of the distribution $f^{(0)}$ from its Maxwellian form, a Sonine polynomial expansion was considered. As usual, for practical purposes, we retained only the first non-zeroth order contribution to the expansion and derived an explicit expression for the second Sonine coefficient $a_2$.

We want in the present work to focus in the properties of the homogeneous steady state, describing in more detail the features of the velocity distribution function. More specifically, our aim here is two-fold. First, as noted in our previous work \cite{GCV13}, we \emph{assume} that in the steady state the homogeneous distribution function $f_\text{s}$ admits a \emph{scaling} solution
\begin{equation}
\label{1.1}
f_\text{s}\to n v_{0,\text{s}}^{-d} \varphi(c,\xi^*),
\end{equation}
where $n$ is the number density, $v_{0,\text{s}}=\sqrt{2T_\text{s}/m}$ is the thermal velocity ($T_\text{s}$ being the steady granular temperature), $\mathbf{c}\equiv \mathbf{v}/v_{0,\text{s}}$ is a dimensionless velocity and $\xi^*$ (defined below in Eq.\ \eqref{2.13}) is the (dimensionless) noise strength of the stochastic term of thermostat. According to the scaling form \eqref{1.1} and in contrast to the results obtained in the homogeneous cooling state (undriven gas) \cite{NE98}, the dependence of the reduced distribution $\varphi$ on temperature is encoded not only through the (reduced) velocity $c$ but also through the driven parameter $\xi^*$. In this paper, we perform Monte Carlo simulations \cite{B94} of the Enskog-Boltzmann equation to confirm that indeed the scaled distribution $\varphi$ presents this universal character for arbitrary values of the coefficient of restitution $\al$ and the external driven parameters. As a second goal, we shall characterize the behavior of $\varphi(c,\xi^*)$ in the domain of thermal velocities by evaluating the two first non-trivial coefficients ($a_2$ and $a_3$) of an expansion of $\varphi$ in Sonine polynomials. Given that both coefficients cannot be exactly obtained, we will propose two different approximations to estimate $a_2$ and $a_3$. In particular, we provide expressions for the coefficient $a_2$ in a more accurate calculation method than in our previous work \cite{GCV13}. Therefore, we give an analytical expression for the distribution function with one more term, and in a more refined approximation.  As we will see, the comparison with the direct simulation Monte Carlo (DSMC) results obtained specifically for this work shows that the analytical expression of the distribution function derived here describes very well the system in a wide range of velocities. A preliminary report of part of the results offered in this paper has been published elsewhere \cite{M12}.

The plan of the paper is as follows. In section \ref{sec2} we describe the system, the thermostat and the system kinetic equation. Next, in section \ref{sec3} we obtain explicit expressions for the two first Sonine coefficients $a_2$ and $a_3$ while the numerical solution of the Enskog-Boltzmann equation for the system studied here is presented in section \ref{comparison} for disks and spheres. A comparison with the approximated theoretical expressions derived in section \ref{sec3} is also carried out, showing in general good agreement between theory and simulation. The paper is closed in section \ref{discussion} with some concluding remarks.


\section{Enskog-Boltzmann kinetic theory for homogeneous driven states}
\label{sec2}


Let us then consider a set of identical smooth hard disks/spheres ($d$ is the dimension of the system) with mass $m$ and diameter $\sigma$ that collide inelastically. At moderate densities, one can still assume that there are no correlations between the velocities of two particles that are about to collide (molecular chaos hypothesis) \cite{GS95}, so that the two-body distribution function factorizes into the product of the one-particle velocity distribution functions $f(\mathbf{r}, \mathbf{v}, t)$. For a spatially uniform state, the Enskog kinetic equation for $f(v, t)$ reads
\begin{equation}
\partial_{t}f+{\cal F}f=\chi J[f,f], \label{2.1}
\end{equation}
where $J[f,f]$ is the collision operator, given by
\begin{equation}
\label{2.2}
J\left[\mathbf{v}_1|f(t), f(t)\right] =\sigma^{d-1}\int
d\mathbf{v}_{2}\int d\widehat{\boldsymbol {\sigma}}\Theta
(\widehat{\boldsymbol {\sigma}}\cdot \mathbf{g}_{12})(\widehat{
\boldsymbol {\sigma }}\cdot \mathbf{g}_{12})\left[ \alpha^{-2}f(\mathbf{v}_{1}^{\prime
})f(\mathbf{v}_{2}^{\prime})-f(\mathbf{v}_{1})f(\mathbf{v}_{2})\right].
\end{equation}
Here, $\cal{F}$ is an operator representing the effect of an external force, $\chi$ is the pair correlation function at contact \cite{CS69}, $\widehat{\boldsymbol {\sigma}}$ is a unit vector along the line joining the centers of the colliding spheres, $\Theta $ is the Heaviside step function and ${\bf g}_{12}={\bf v}_{1}-{\bf v}_{2}$ is the relative velocity. In addition, the primes on the velocities in equation \ \eqref{2.2} denote the initial values $\{\mathbf{v}_1', \mathbf{v}_2'\}$ that lead to $\{\mathbf{v}_1, \mathbf{v}_2\}$ following a binary collision:
\begin{equation}
\label{2.3}
\mathbf{v}_{1}'=\mathbf{v}_{1}-\frac{1}{2}(1+\alpha^{-1})(\widehat{{\boldsymbol {\sigma }}}\cdot
\mathbf{g}_{12})\widehat{\boldsymbol {\sigma}}, \quad \mathbf{v}_{2}'=\mathbf{v}_{2}+\frac{1}{2}(1+\alpha^{-1})(\widehat{{\boldsymbol {\sigma }}}\cdot
\mathbf{g}_{12})\widehat{\boldsymbol {\sigma}},
\end{equation}
where $\alpha \leq 1$ is the (constant) coefficient of normal restitution. Except for the presence
of the factor $\chi$ (which accounts for the increase of the collision frequency due to excluded volume effects), the Enskog equation for \emph{uniform} states is identical to the Boltzmann equation for a low-density gas. For this reason, henceforth we will call the Enskog-Boltzmann equation to Eq.\ \eqref{2.1}.





As we said in the Introduction, our granular gas is subjected to homogeneous volume forces that try to mimic the interaction with a surrounding molecular fluid \cite{Puglisi}. These forces, usually called ``thermostats'' \cite{E90},  show up in the kinetic equation \eqref{2.1} through the term $\mathcal{F}$. Here, we will consider a volume force composed by two independent terms. One term corresponds to a Gaussian white noise force ($\mathbf{F}^\text{st}$), that tries to simulate the kinetic energy gain due to eventual collisions with the (more rapid) molecules of the surrounding fluid. It does this by adding a ``random'' velocity to each particle. This additional velocity is extracted from a Maxwellian distribution with a characteristic variance determined by the
``noise intensity'' $\xi_b^2$ \cite{WM96}. The other term corresponds to a drag force ($\mathbf{F}^\text{drag}$)
of the form $-\gamma_b\mathbf{v}_i(t)$, that tries to capture the effect of the surrounding fluid
viscosity ($\gamma_b$ is a drag coefficient). This kind of thermostat composed by two different forces has been used by other authors in previous works \cite{Puglisi}.
The total thermostat force $\mathbf{F}^\text{th}(t)$ is
\begin{equation}
\label{2.5}
{\bf F}_i^{\text{th}}(t)={\bf F}_i^{\text{st}}(t)+{\bf F}_i^{\text{drag}}(t)=
{\bf F}_i^{\text{st}}(t)-\gamma_\text{b}{\bf v}_i(t).
\end{equation}
Since ${\bf F}_i^{\text{st}}(t)$ is a Gaussian white noise \cite{WM96}, it fulfills the conditions \cite{MS00}
\begin{equation}
\label{2.6}
\langle {\bf F}_i^{\text{st}}(t) \rangle ={\bf 0}, \quad
\langle {\bf F}_i^{\text{st}}(t) {\bf F}_j^{\text{st}}(t') \rangle =\mathbf{1}m^2 \xi_\text{b}^2 \delta_{ij}\delta(t-t'),
\end{equation}
where $\mathbf{1}$ is the $d\times d$ unit matrix and $\delta_{ij}$ is the Kronecker delta function. The corresponding term in the Enskog-Boltzmann equation associated to the stochastic force ${\bf F}_i^{\text{st}}$ is represented by the Fokker-Planck operator $-\frac{1}{2}\xi_\text{b}^2\partial^2/\partial v^2$ \cite{NE98}. Therefore, the stochastic and drag forces contribute to the kinetic equation with terms of the form
\begin{equation}
  \label{calF}
  \mathcal{F}f=\mathcal{F}^{\text{st}}f+\mathcal{F}^{\text{drag}}f, \quad
  \mathcal{F}^{\text{st}}f=-\frac{1}{2}\xi_\text{b}^2\frac{\partial^2}{\partial v^2} f, \quad
  \mathcal{F}^{\text{drag}}f=-\frac{\gamma_\text{b}}{m}\frac{\partial}{\partial\mathbf{v}}\cdot\mathbf{v}f.
\end{equation}
Notice that the thermostat terms  $\mathcal{F}^\text{st}$ and $\mathcal{F}^\text{drag}$ introduce in the kinetic equation \eqref{2.1} two new and independent time scales given by $\tau_\text{st}=v_0^2/\xi_\text{b}$ and $\tau_\text{drag}=m/\gamma_\text{b}$, respectively. Here, $v_0=\sqrt{2T/m}$ is the thermal velocity defined in terms of the granular temperature $T$. A similar external driving force to that of equation \eqref{2.6} has been recently proposed to model the effect of the interstitial fluid on grains in monodisperse gas-solid suspensions \cite{GTSH12}.

The Enskog-Boltzmann equation \eqref{2.1} can be more explicitly written when one takes into account the form \eqref{calF} of the forcing term $\mathcal{F}f$. It is given by
\begin{equation}
\partial_{t}f-\frac{\gamma_\text{b}}{m}
\frac{\partial}{\partial{\bf v}}\cdot {\bf v}
f-\frac{1}{2}\xi_\text{b}^2\frac{\partial^2}{\partial v^2}f=\chi J[f,f]. \label{2.7}
\end{equation}
The density $n$ and temperature $T$ fields are defined as usual (except that for our case the mean flow velocity vanishes)
\begin{equation}
  \label{n}
  n(t)=\int\;\dd\mathbf{v}f(\mathbf{v},t),
\end{equation}
\begin{equation}
\label{2.8}
T(t)=\frac{m}{d n}\; \int\; \dd \mathbf{v}\; v^2 f(\mathbf{v},t).
\end{equation}
The balance equation for the homogeneous temperature can be easily obtained by multiplying both sides of equation \ \eqref{2.7} by $v^2$ and integrating over velocity. The result is
\begin{equation}
\label{Tdt}
\partial_tT=-\frac{2T}{m}\gamma_\text{b} +m\xi_\text{b}^2-
\zeta T,
\end{equation}
where
\begin{equation}
\label{zeta}
\zeta=-\frac{m}{dnT}\int\; \dd\mathbf{v}\;  v^2 J[f,f]
\end{equation}
is the cooling rate $\zeta$. It is proportional to $1-\alpha^2$ and is due to the inelastic character of the collisions.

We will assume now that a \emph{normal} (or hydrodynamic) solution to equation\ \eqref{2.7} exists. This means that all the time dependence of the distribution function $f(v,t)$ occurs through a functional dependence on the hydrodynamic fields \cite{CC70,McL89}. Since  the temperature is the only relevant field in this problem, the time dependence of $f(v,t)$ occurs only through $T(t)$. Therefore, according to equation\ \eqref{Tdt}, one gets
\begin{equation}
\label{fhd}
\partial_t f=\frac{\partial f}{\partial T}\partial_tT= -\left(\frac{2\gamma_\text{b}}{m} -\frac{m}{T}\xi_\text{b}^2+
\zeta\right)T \frac{\partial f}{\partial T}.
\end{equation}
Substitution of equation\ (\ref{fhd}) into equation\ (\ref{2.7}) yields
\begin{equation}
\label{2.12} -\left( \frac{2}{m}\gamma_\text{b}-\frac{m}{T}\xi_\text{b}^2+ \zeta\right)T
\frac{\partial f}{\partial T}-\frac{\gamma_\text{b}}{m} \frac{\partial}{\partial {\bf
v}}\cdot {\bf v} f-\frac{1}{2}\xi_\text{b}^2\frac{\partial^2}{\partial v^2}f=\chi J[f,f].
\end{equation}

After a transient regime, it is expected that the gas reaches a steady state characterized by a constant granular temperature $T_\text{s}$. In this case, $\partial_t T=0$ and the balance equation \eqref{Tdt} leads to
\begin{equation}
\label{2.9}
T_\text{s}=\frac{m\xi_\text{b}^2}{\zeta_\text{s}+\frac{2\gamma_\text{b}}{m}},
\end{equation}
where the subindex $\text{s}$ means that the quantities are evaluated in the steady state. Given that equation \eqref{2.9} establishes a relation between the two driven parameters $\gamma_\text{b}$ and $\xi_\text{b}^2$, only one of the above parameters will be considered as independent. Henceforth, we will take $\xi_\text{b}^2$ as the relevant driven parameter. For elastic collisions ($\al=1$), $\zeta_\text{s}=0$ and so, equation \eqref{2.9} yields $T_\text{s}=T_\text{b}$ where
\beq
\label{2.10.1}
T_\text{b}=\frac{m^2\xi_\text{b}^2}{2\gamma_\text{b}}.
\eeq
As in the work of Gradenigo et al. \cite{GSVP11}, equation \eqref{2.10.1} defines a ``bath temperature''. Its name may be justified since it is determined by the two thermostats parameters ($\gamma_\text{b}$ and $\xi_\text{b}^2$), and thus it can be considered as remnant of the physical temperature of the surrounding ordinary (elastic) fluid. In this sense,  for elastic collisions, $T=T_\text{b}$, and so energy equipartition is fulfilled (in accordance to equilibrium statistical mechanics principles). For inelastic gases (i.e., for $\alpha<1$, $\zeta_\text{s}>0$), equation \eqref{2.9} yields $T<T_\text{b}$. From a physical point of view, it makes sense that the inelastic granular gas is cooler than the surrounding ordinary fluid.

The kinetic equation of the steady distribution function $f_s$ can be easily obtained by using the relation \eqref{2.9} in equation \eqref{2.12}:
\begin{equation}
\label{2.11}
\frac{1}{2}\zeta_\text{s} \frac{\partial}{\partial {\bf
v}}\cdot {\bf v} f_\text{s}-\frac{m\xi_\text{b}^2}{2T_\text{s}} \frac{\partial}{\partial {\bf
v}}\cdot {\bf v} f_\text{s}-\frac{1}{2}\xi_\text{b}^2\frac{\partial^2}{\partial v^2}f_\text{s}=\chi J[f_\text{s},f_\text{s}].
\end{equation}
Equation \eqref{2.11} clearly shows that $f_\text{s}(v)$ must also depend on the model parameter $\xi_\text{b}^2$  and the steady temperature $T_\text{s}$ apart from its dependence on the coefficient of restitution $\al$. Note that the steady cooling rate $\zeta_\text{s}$ is defined in terms of the steady distribution $f_s$ (see equation \eqref{zeta}).

Based on previous results obtained for undriven \cite{NE98,GS95} and driven \cite{NE98,MS00,GCV13} systems, it is expected that equation\ \eqref{2.11} admits a scaling solution of the form given by equation \eqref{1.1},
where $\varphi$ is an unknown function of the dimensionless parameters
\begin{equation}
\label{2.13}
\mathbf{c}\equiv \frac{\mathbf{v}}{v_{0,\text{s}}}, \quad \xi^*\equiv \frac{m\ell}{\chi T_\text{s}v_{0,\text{s}}}\xi_\text{b}^2.
\end{equation}
Here,
\begin{equation}
  \label{mfp}
  \ell=\frac{1}{n\sigma^{d-1}}
\end{equation}
is the mean free path for hard spheres. In the steady state, it is also convenient to define  the collision frequency
\begin{equation}
  \label{cf}
  \nu_s=\frac{v_{0,\text{s}}}{\ell}=\sqrt{\frac{2T_\text{s}}{m}}n\sigma^{d-1},
\end{equation}
and the reduced drag coefficient
\begin{equation}
  \label{reddrag}
  \gamma^*=\frac{\gamma_\text{b}}{\chi m\nu_s}.
\end{equation}
In terms of the (reduced) distribution function $\varphi$, equation\ \eqref{2.11} may be rewritten as
\begin{equation}
\label{2.15} \frac{1}{2}\zeta^* \frac{\partial}{\partial {\bf
c}}\cdot {\bf c}\varphi-\frac{1}{2}\xi^*\frac{\partial}{\partial {\bf c}}\cdot {\bf c} \varphi- \frac{1}{4}\xi^* \frac{\partial^2}{\partial c^2}\varphi= J^*[\varphi,\varphi],
\end{equation}
where we have introduced the dimensionless quantities
\begin{equation}
\label{2.16}
\zeta^*\equiv\frac{\zeta_\text{s}}{\chi\nu_s}, \quad J^*[\varphi,\varphi]
\equiv \frac{ v_{0,\text{s}}^{d}}{n\nu_s}J[f,f].
\end{equation}
Equation \eqref{2.15} clearly shows that the dependence of the scaled distribution function $\varphi$ on the temperature is encoded through two \emph{different} parameters: the dimensionless velocity $c$ and the (reduced) noise strength $\xi^*$. This scaling differs from the one assumed in the free cooling case \cite{NE98} where only the dimensionless velocity $c$ is required to characterize the distribution $\varphi$.  A similar scaling solution to the form \eqref{2.12} has been recently found \cite{GMT12} at all times (also for unsteady states) in the particular case $\gamma_\text{b}=0$. Thus, our guess in equation \eqref{1.1} seems to be reasonable.

In the case of elastic particles ($\al=1$), the cooling rate $\zeta_\text{s}^*$ vanishes and the solution of equation\ \eqref{2.15} is the Maxwellian distribution
\begin{equation}
\label{2.17}
\varphi_\text{M}(c)= \pi^{-d/2}e^{-c^2}.
\end{equation}
However, if the particles collide inelastically ($\alpha <1$), then $\zeta^*\neq 0$ and the exact form of $\varphi(c)$ is not known. One of the objectives of the present work is to find an accurate analytical solution to the reduced distribution function $\varphi$, from equation \eqref{2.15}. As usual, the behavior of $\varphi(c,\xi^*)$ in the region of thermal velocities ($c\simeq 1$) can be well characterized by the two first nontrivial coefficients ($a_2$ and $a_3$) of an expansion in Sonine polynomials. This will be done in the next section by considering two different approaches.

In reduced units, the steady-state condition \eqref{2.9} can be simply written as
\begin{equation}
\label{ss}
2\gamma^*=\xi^*-\zeta^*.
\end{equation}
Since $\gamma^* \geq 0$, then equation \eqref{ss} requires $\xi^*\geq \zeta^*$. Thus, at a given value of the coefficient of restitution $\al$, there is a minimum threshold value $\xi^*_\text{th}(\al)$  of the noise intensity needed to reach a steady state. The value of $\xi^*_\text{th}$ coincides with the (reduced) cooling rate $\zeta^*(\al)$. Given that the latter cannot be exactly determined, a good estimate of it is obtained when one replaces the true distribution $\varphi$ by its Maxwellian form $\varphi_\text{M}$. In this case, $\zeta^*\to \zeta_\text{M}^*$, where \cite{GS95}
\beq
\label{ss.1}
\zeta_\text{M}^*=\frac{\sqrt{2}}{d}\frac{\pi^{(d-1)/2}}{\Gamma(d/2)}(1-\al^2).
\eeq

Before closing this section, let us make some observations. First, due to the equivalence between the Enskog and Boltzmann equations in the homogeneous states, the solution to equation \eqref{2.15} does not depend explicitly on the pair correlation function $\chi$ (and $\chi$ is a function of the packing fraction). This means that the reduced distribution function $\varphi$ has the same universal form for arbitrary values of the packing fraction of the granular fluid. Thus, we do not need to provide explicit expressions for the purpose of this work, although they may be found elsewhere \cite{CS69,T95}. Also, as mentioned before, the steady-state equation \eqref{2.9} leads to a relation between $\xi_\text{b}^2$ and $\gamma_\text{b}$ so that the scaled distribution $\varphi$ depends on both parameters only through the reduced noise strength $\xi^*$.
Therefore, providing $\xi^*$ (and not $\xi_\text{b}^2$, $\gamma_\text{b}$, nor $\chi$) is enough to determine uniquely the steady distribution function. We will check that in effect this is the case in section \ref{comparison}, from comparison with simulation data.

\section{Analytical solution of the scaled distribution function}
\label{sec3}

The goal of this section is to determine a perturbative (although sufficiently accurate) analytic solution of the distribution function $\varphi(c,\xi^*)$. As said before, a convenient and useful way of characterizing $\varphi(c,\xi^*)$ in the range of low to intermediate velocities is through the Sonine polynomial expansion
\begin{equation}
\label{3.1}
\varphi(c,\xi^*)=\varphi_\text{M}(c)\left[1+\sum_{p=1}^\infty\; a_p(\xi^*)\; S_p(c^2)\right],
\end{equation}
where $S_p$ are generalized Laguerre or Sonine polynomials. They are defined as \cite{BP04}
\beq
\label{3.2}
S_p(x)=\sum_{k=0}^p\;\frac{(-1)^k \left(\frac{d}{2}-1+p\right)!}{\left(\frac{d}{2}-1+k\right)!(p-k)!k!}x^k,
\eeq
and satisfy the orthogonality relations \cite{Abramowitz}
\beq
\label{3.6}
\int\; \dd \mathbf{c}\; \varphi_\text{M}(c)\; S_p(c^2)\; S_{p'}(c^2)={\cal N}_p\;\delta_{p p'},
\eeq
where ${\cal N}_p$ is a normalization constant.

The first few Sonine polynomials relevant for our study are
\beq
\label{3.4}
S_0(x)=1, \quad S_1(x)=-x+\frac{d}{2}, \quad S_2(x)=\frac{1}{2}x^2-\frac{d+2}{2}x+\frac{d(d+2)}{8},
\eeq
\beq
\label{3.5}
S_3(x)=-\frac{1}{6}x^3+\frac{d+4}{4}x^2-\frac{(d+2)(d+4)}{8}x+\frac{d(d+2)(d+4)}{48}.
\eeq
The coefficients $a_p$ appearing in equation \eqref{3.1} (also called \textit{cumulants}) are the corresponding velocity moments of the scaling function $\varphi$, i.e.,
\beq
\label{3.6_31}
a_p(\xi^*)=\frac{1}{{\cal N}_p}\int\; \dd \mathbf{c}\; S_p(c^2)\; \varphi(c,\xi^*).
\eeq
In particular, the temperature definition \eqref{2.8} implies $\langle c^2 \rangle =\frac{d}{2}$ and therefore,
\beq
\label{3.7}
a_1=\frac{2}{d}\langle S_1(c^2) \rangle=0.
\eeq
Here, $\langle \cdots \rangle$ denotes an average over the scaled distribution $\varphi$, namely,
\beq
\label{3.8}
\langle c^p \rangle\equiv \int\; \dd \mathbf{c}\; c^p\; \varphi(c).
\eeq
In the present work, we will retain up to the first two nontrivial coefficients $a_2$ and $a_3$. They are related to the fourth and sixth velocity moments as
\begin{equation}
\label{3.9}
\langle c^4 \rangle =\frac{d(d+2)}{4}(1+a_2),
\end{equation}
\begin{equation}
\label{3.10}
\langle c^6 \rangle =\frac{d(d+2)(d+4)}{8}(1+3a_2-a_3).
\end{equation}

In order to determine the coefficients $a_k$, we construct a set of equations for the velocity moments $\langle c^{2p} \rangle$. The hierarchy for the moments can be easily derived by multiplying both sides of equation\ \eqref{2.15} by $c^{2p}$ and integrating over $\mathbf{c}$. The result is
\begin{equation}
\label{3.11}
p(\zeta^*-\xi^*)\langle c^{2p} \rangle+\frac{p(2p+d-2)}{2}\xi^*\langle c^{2p-2} \rangle=\mu_{2p},
\end{equation}
where
\begin{equation}
\label{3.12}
\mu_{2p}=-\int\; d{\bf c}\;c^{2p}\; J^*[\varphi,\varphi].
\end{equation}
Upon writing equation\ \eqref{3.11} use has been made of the results
\begin{equation}
\label{3.13}
\int\; d{\bf c}\; c^{2p}\; \frac{\partial}{\partial {\bf
c}}\cdot {\bf c}\varphi(\mathbf{c})=-2p\langle c^{2p} \rangle,
\end{equation}
\begin{equation}
\label{3.14}
\int\; \dd {\bf c}\; c^{2p}\; \frac{\partial^2}{\partial c^2}\varphi(\mathbf{c})=2p(2p+d-2)\langle c^{2p-2} \rangle.
\end{equation}
Note that, according to equations \eqref{zeta} and \eqref{3.12}, the (reduced) cooling rate $\zeta^*=\frac{2}{d}\mu_2$.

The cumulants $a_p$ can be obtained from the \emph{exact} set of moment equations \eqref{3.11}. However, given that the collisional moments $\mu_{2p}$ are functionals of the distribution $\varphi$, equation\ \eqref{3.11} becomes an infinite hierarchy of moment equations. In other words, \emph{all} the Sonine coefficients $a_p$ are coupled and so, one has to resort to some kind of truncation in the series \eqref{3.1} to get explicit forms for $a_p$. Thus, based on the expectation that the Sonine coefficients are small, one usually \emph{approximates} the first few collisional moments $\mu_{2p}$ by inserting the expansion \eqref{3.1} into equation\ \eqref{3.12}, truncating the expansion in a given order and, in some cases, neglecting nonlinear terms. In particular, in the case of the collisional moments defined by equation\ \eqref{3.12} with $p=1$, 2, and 3, one gets
\begin{equation}
\label{3.15}
\mu_2\to A_0+A_2 a_2+A_3 a_3,
\end{equation}
\begin{equation}
\label{3.16}
\mu_4\to B_0+B_2 a_2+B_3 a_3,
\end{equation}
\begin{equation}
\label{3.17}
\mu_6\to C_0+C_2 a_2+C_3 a_3.
\end{equation}
The expressions of the coefficients $A_i$, $B_i$, and $C_i$ as functions of the coefficient of restitution $\alpha$ and the dimensionality $d$ were independently derived by van Noije and Ernst \cite{NE98} and by Brilliantov and P\"oschel \cite{BP06}. They are displayed in the Appendix \ref{appA} for the sake of completeness. We need to note that in equations \eqref{3.15}--\eqref{3.17}, we have neglected the coefficients $a_p$ with $p\geq 4$ and nonlinear terms (like $a_2^2$, $a_2 a_3$, and $a_3^2$).

The exact moment equation \eqref{3.11} becomes an approximation when it is linearized with respect to $a_2$ and $a_3$. For $p=2$, one gets
\begin{equation}
\label{3.18}
\left[B_2-(d+2)(A_0+A_2)+\frac{d(d+2)}{2}\xi^*\right]a_2+\left[B_3-(d+2)A_3\right]a_3=(d+2)A_0-B_0,
\end{equation}
while the result for $p=3$ is
\begin{equation}
\label{3.19}
\left[C_2+\frac{3}{4}(d+2)(d+4)(d\xi^*-3A_0-A_2)\right]a_2+\left[C_3-\frac{3}{4}(d+2)(d+4)
\left(A_3-A_0+\frac{d}{2}\xi^*\right)\right]a_3=\frac{3}{4}(d+2)(d+4)A_0-C_0.
\end{equation}
The (reduced) thermostat parameter $\xi^*$ depends on the value of the (steady) granular temperature, which is a function of $a_2$ and $a_3$. Since it is expected that both coefficients are quite small, we evaluate $\xi^*$ by assuming $a_2=a_3=0$. In this case, the set of Eqs.\ \eqref{3.18} and \eqref{3.19} becomes a simple linear algebraic set of equations that can be easily solved to give $a_2$ and $a_3$ in terms of $d$, $\al$ and $\xi^*$. As noted previously by Montanero and Santos \cite{MS00,SM09}, there is a certain degree of ambiguity in the approximations used in the determination of $a_2$ and $a_3$. Here, in order to solve the set of equations \eqref{3.18} and \eqref{3.19}, we consider two basic classes of approximations. \vicente{ In Approximation I, we first assume that $a_3\ll a_2$ so that, $a_3$ can be neglected versus $a_2$ in Eq.\ \eqref{3.18} but not in Eq.\ \eqref{3.19}. This is equivalent to neglect $a_3$ in Eqs.\  \eqref{3.15} and \eqref{3.16} for $\mu_2$ and $\mu_4$, respectively. Given that $\mu_6$ is expected to be smaller than $\mu_4$, it seems to be more accurate to neglect $a_3$ in Eq.\ \eqref{3.18} rather than in Eq.\ \eqref{3.19}. The comparison with computer simulations confirm this expectation.} In Approximation II, both Sonine coefficients $a_2$ and $a_3$ are considered as being of the same order of magnitude. Since the latter does not assume negligible contributions of $a_3$ to the expression of $a_2$, this approximation should be more accurate.

In Approximation I, the expression of the second Sonine coefficient $a_2$ may be calculated independently of $a_3$, from equation \eqref{3.18} with $a_3=0$. In fact, $a_2^{(I)}$ was obtained in previous works \cite{M12,GCV13}. Its explicit expression is given by equation \eqref{b2} while $a_3^{(I)}$ is
\begin{equation}
\label{b3bis}
a_3^{(I)}(\alpha,\xi^*)=F\left(\alpha,a_2^{(I)}(\alpha),\xi^*\right),
\end{equation}
where the function $F(\al, a_2, \xi^*)$ is given by equation \eqref{b4}. The expressions in Approximation II have the following forms
\begin{equation}
\label{b5}
a_2^{(II)}(\alpha,\xi^*)=\frac{M(\alpha,\xi^*)}{N(\alpha,\xi^*)},
\end{equation}
\begin{equation}
\label{b8.1}
a_3^{(II)}(\alpha,\xi^*)=F\left(\alpha,a_2^{(II)}(\alpha),\xi^*\right),
\end{equation}
where the explicit (and rather large) expressions of $M(\alpha,\xi^*)$ and $N(\alpha,\xi^*)$ are given by equations \eqref{b6} and \eqref{b7}, respectively.


\section{Computer simulation results. Comparison with theoretical approaches}
\label{comparison}

In this section, the relevant quantities of the problem ($a_2$, $a_3$ and $\varphi$) will be computed by numerically solving the Enskog-Boltzmann equation by means of the DSMC method \cite{B94,P05}. The DSMC method has proven to be a very efficient tool by solving numerically the Enskog-Boltzmann equation \cite{B94,MS00} for inelastic collisions. The numerical results will be also compared with the theoretical predictions described in section \ref{sec3}. Before doing it, let us provide some details on the implementation of the DSMC method to the problem considered in this paper.

\subsection{Direct simulation Monte Carlo method}

By means of the DSMC method we can obtain a numerical solution of the kinetic equation \eqref{2.7}. This solution has the following advantages: 1) it also determines homogeneous non-steady states; and  2) it does not assume \textit{a priori} nor a \emph{normal} solution neither the specific scaling form \eqref{1.1} of the distribution function, as we did for the analytical solution. Therefore, a comparison of both numerical and analytical solutions is a direct way of validating (for steady states) the hypotheses of existence of a normal solution and of the special scaling form of equation \eqref{2.11}. Also, the comparison will allow us to assess the accuracy of the approximated expressions for $a_2$ and $a_3$. Additionally, it will allow us to show in this work the first preliminary analysis of the transient regime towards the steady state for the kind of thermostat  we are using.

\begin{figure*}
\begin{center}
\begin{tabular}{lr}
\resizebox{7.5cm}{!}{\includegraphics{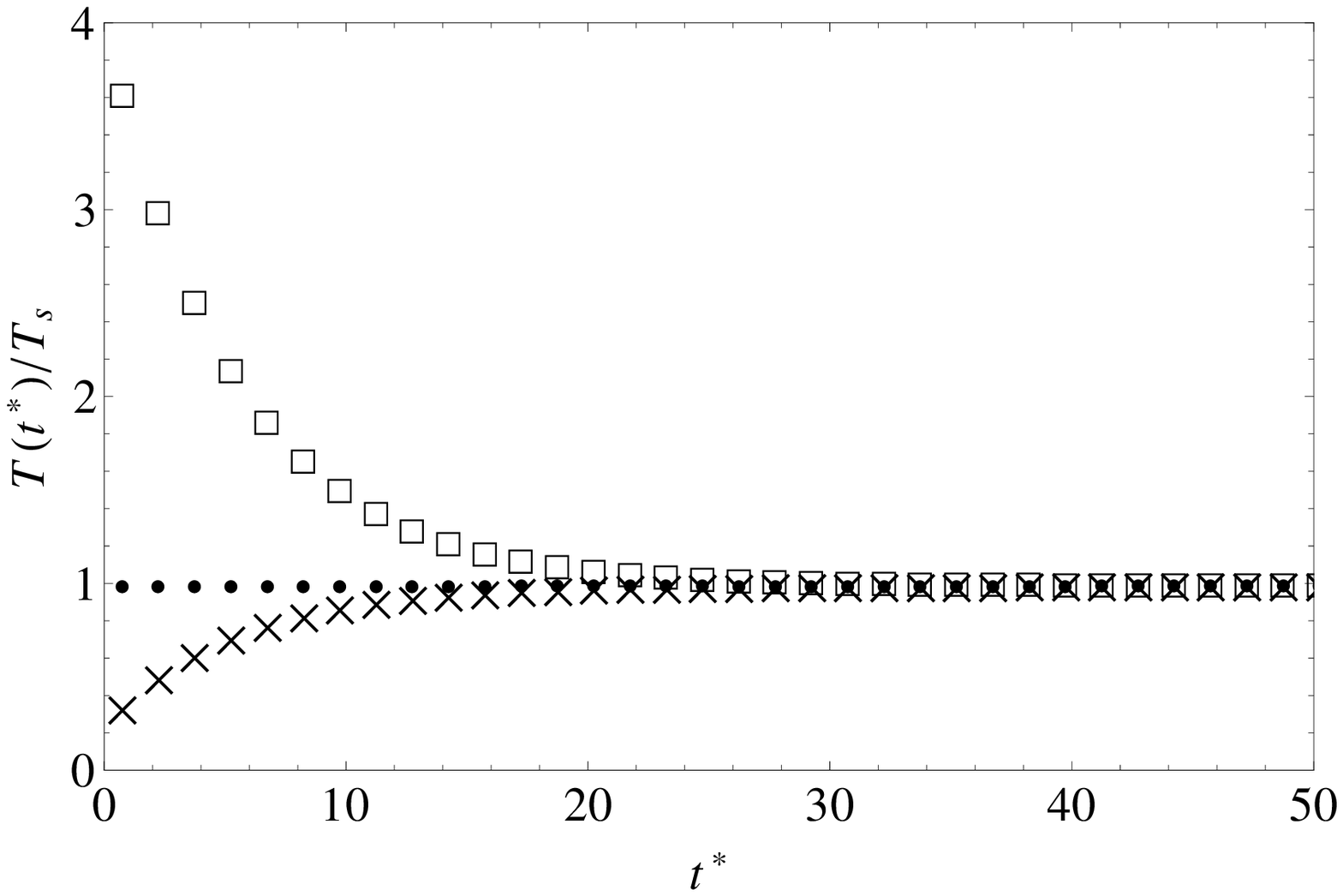}}&\resizebox{7.5cm}{!}
{\includegraphics{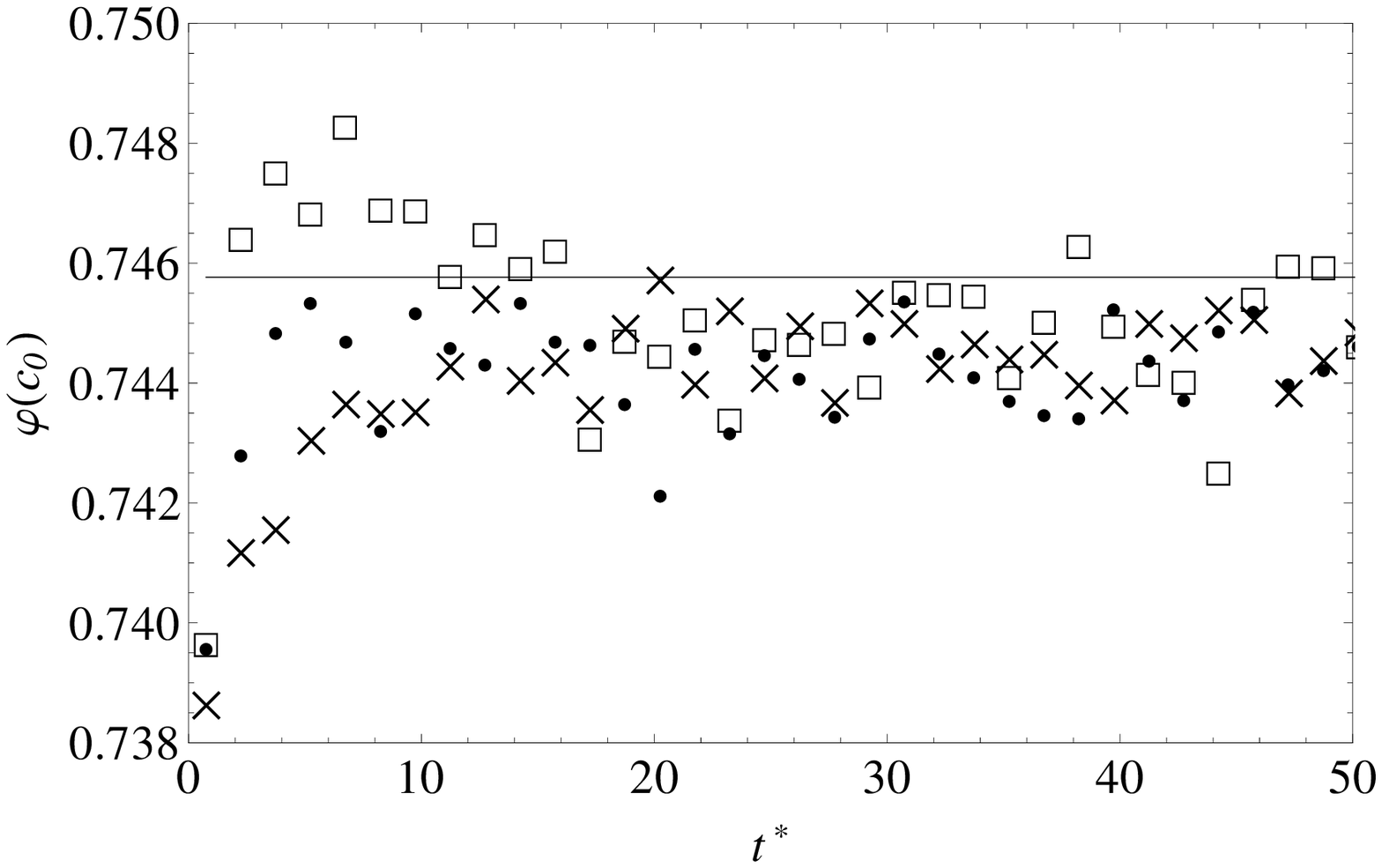}}
\end{tabular}
\end{center}
\caption{Time evolution for hard disks of the reduced temperature $T(t)/T_\text{s}$ (left panel) and the scaled distribution function $\varphi(c_0)$ (right panel) for $\xi^*=0.478$, $\gamma^*=0.014$, and $\alpha=0.8$. Three different initial temperatures have been considered: $T(0)/T_\text{s}=0.25 (\times), 1 (\cdots),$ and $4 (\square)$. Here, $T_\text{s}$ is the steady value of the temperature and $c_0(t)=v_{0,\text{s}}/v_0(t)$, $v_{0,\text{s}}=\sqrt{2T_\text{s}/m}$ being the steady value of thermal speed. The symbols correspond to the simulation results while the horizontal lines refer to the theoretical predictions for $T_\text{s}$ and  $\varphi(c_0)$. The latter has been obtained by retaining the three first Sonine polynomials (see equation \eqref{4.1}) and evaluating  $a_2$ and $a_3$ with Approximation II. Time is measured in units of $\nu^{-1}$ ($t^*=t \nu^{-1})$.
\label{fig1}}
\end{figure*}
\begin{figure*}
\begin{center}
\begin{tabular}{lr}
\resizebox{7.5cm}{!}{\includegraphics{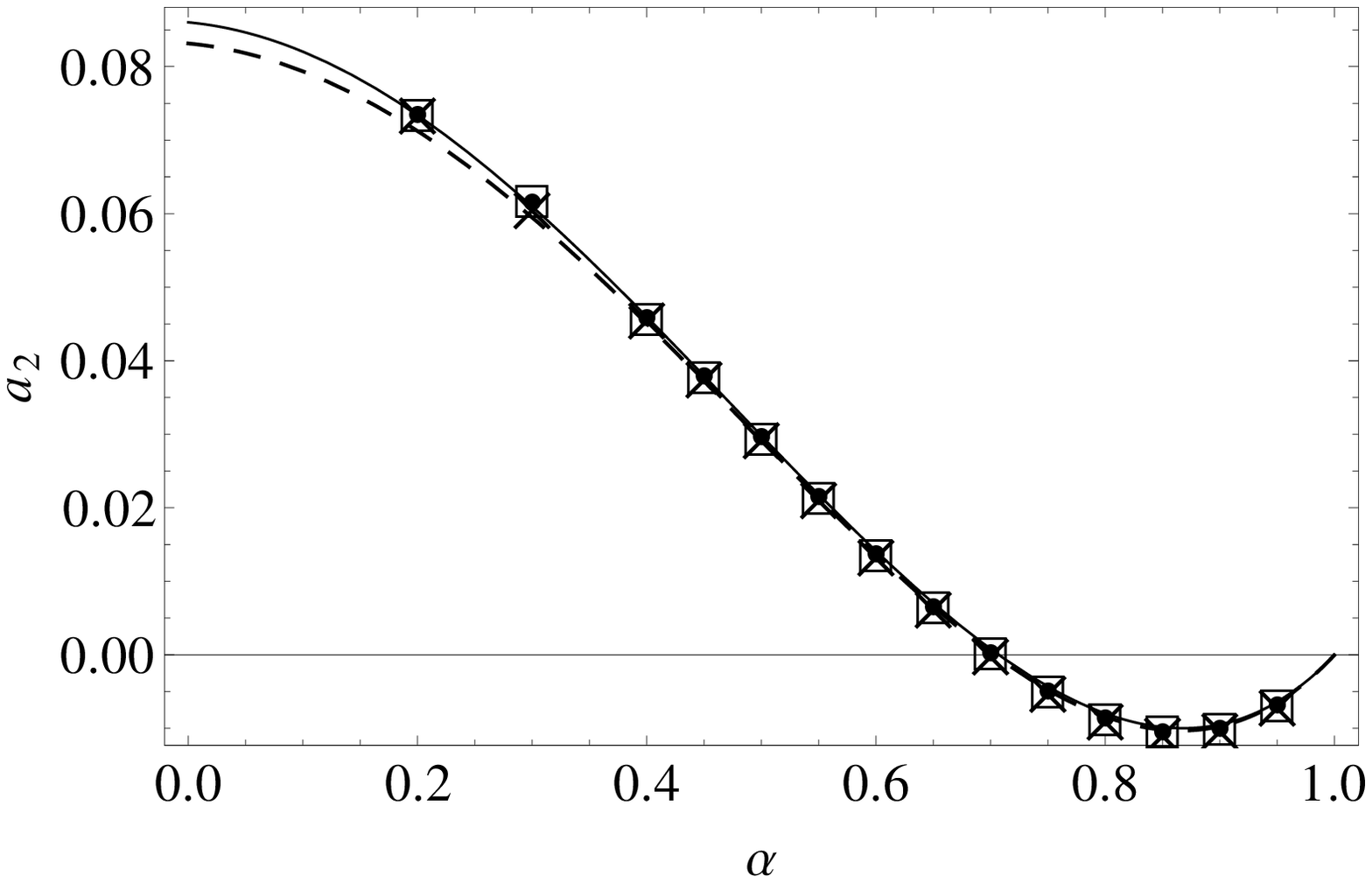}}&\resizebox{7.5cm}{!}
{\includegraphics{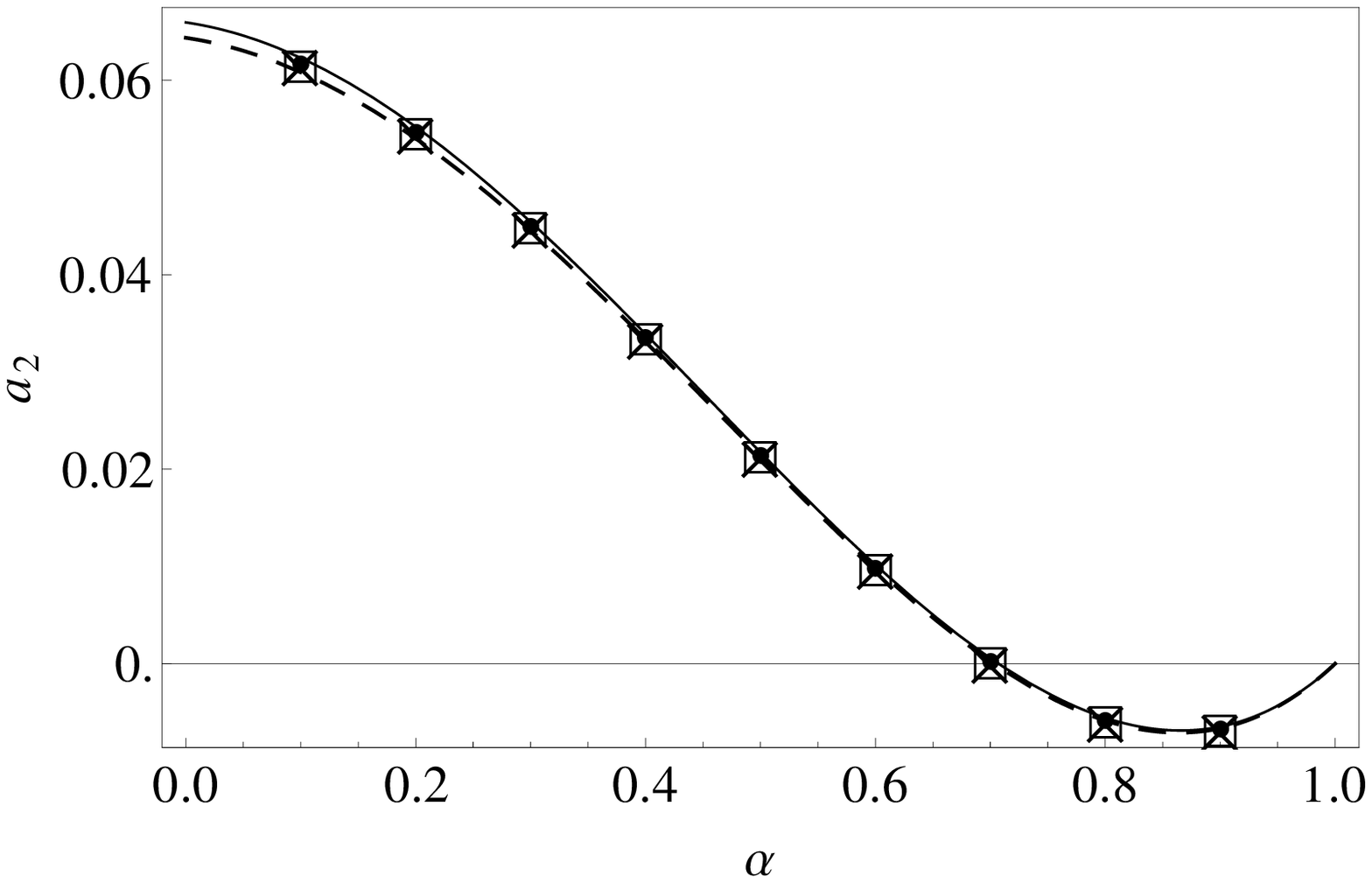}}
\end{tabular}
\end{center}
\caption{Plot of the second Sonine coefficient $a_2$ versus the coefficient of restitution $\al$ for hard disks (left panel) and hard spheres (right panel). The symbols refer to three different systems with different values of the simulation parameters $\gamma_\text{sim}^*$ and $\xi_\text{sim}^*$ but with the same value of $\xi^*$ ($\xi^*=1.26$ for disks and $\xi^*=1.68$ for spheres). The solid and dashed lines are the values obtained for $a_2$ by means of Approximation I and Approximation II, respectively.
\label{fig2}}
\end{figure*}
\begin{figure*}
\begin{center}
\begin{tabular}{lr}
\resizebox{7.5cm}{!}{\includegraphics{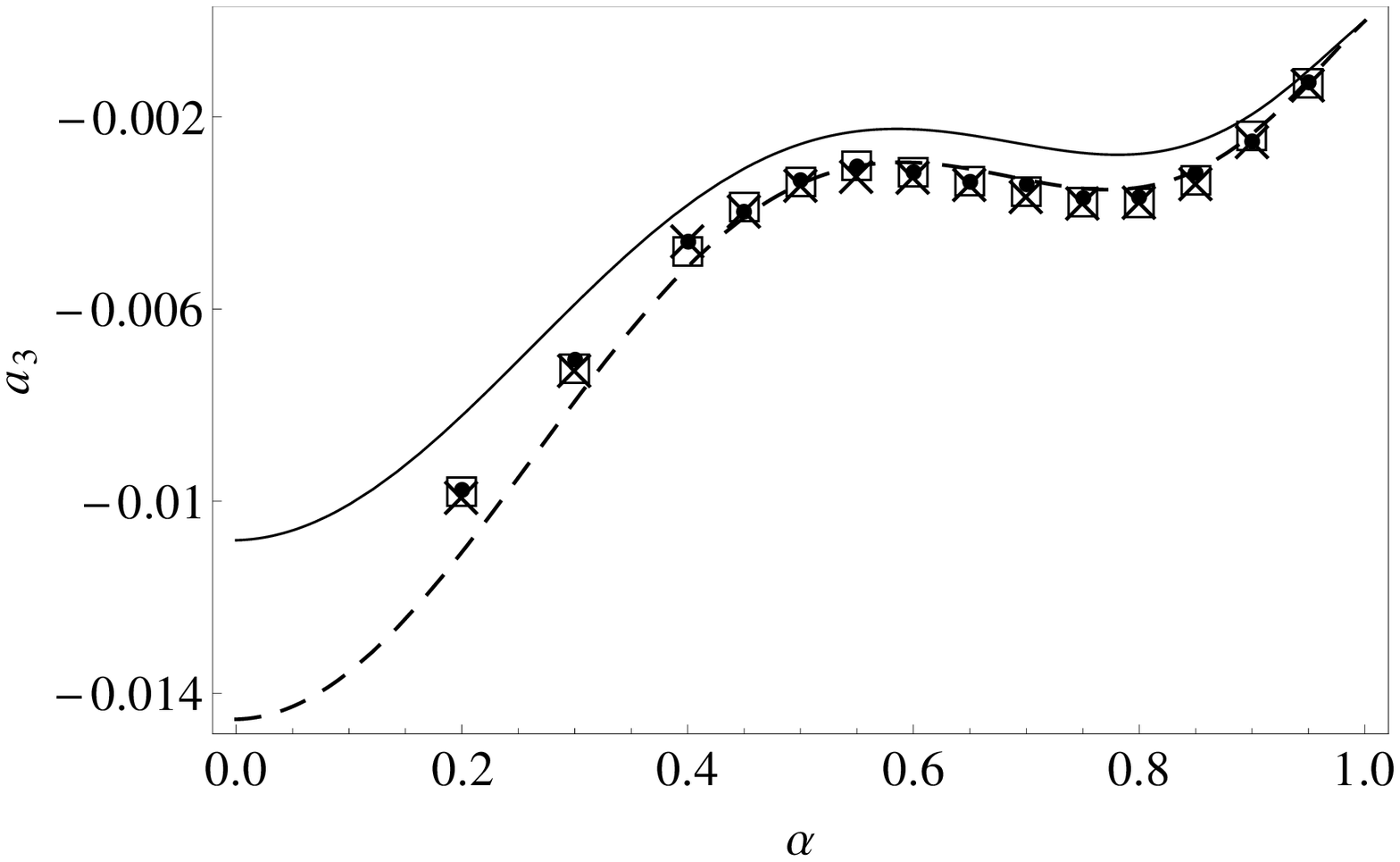}}&\resizebox{7.5cm}{!}
{\includegraphics{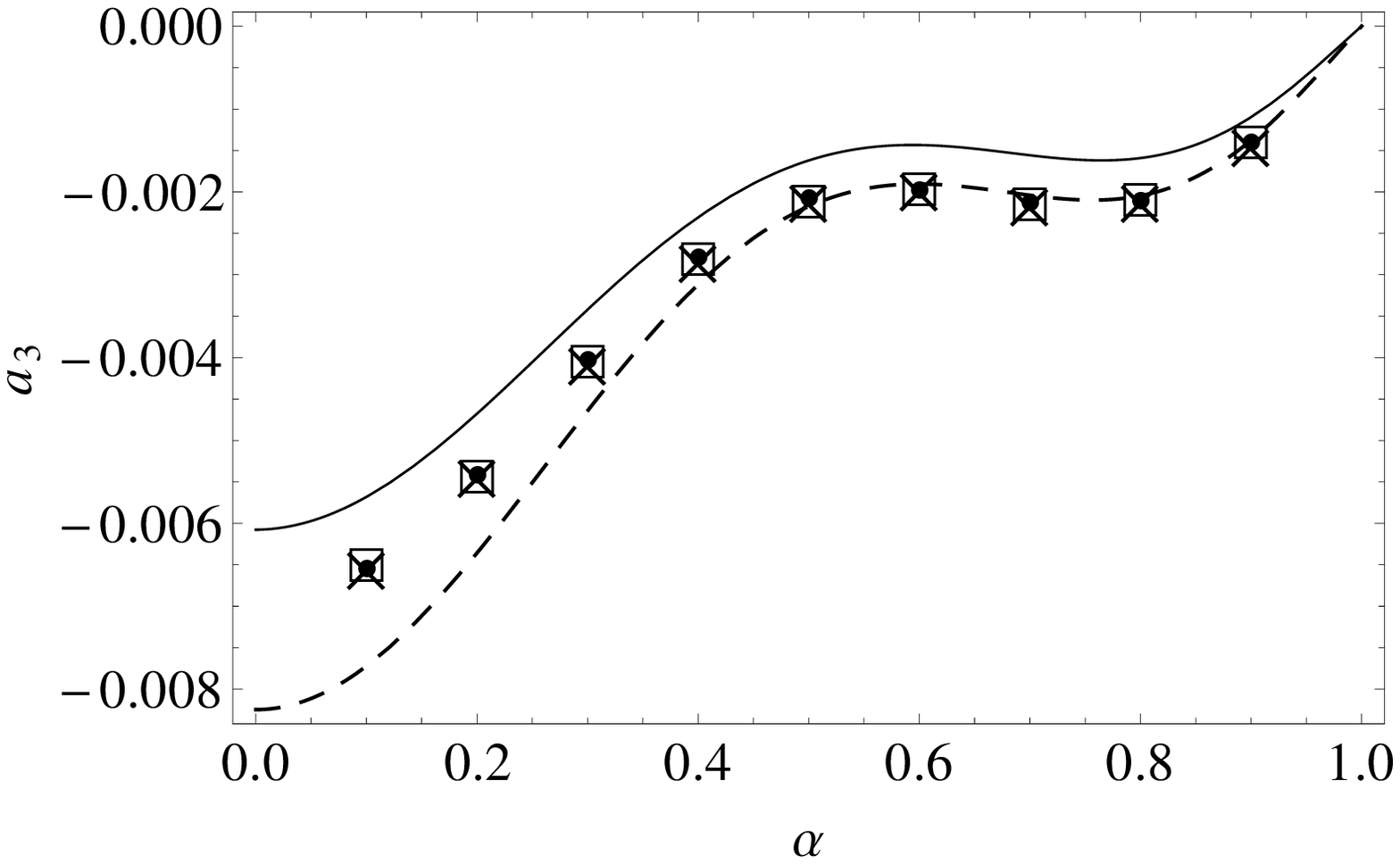}}
\end{tabular}
\end{center}
\caption{Plot of the third Sonine coefficient $a_3$ versus the coefficient of restitution $\al$ for hard disks (left panel) and hard spheres (right panel). The symbols refer to three different systems with different values of the simulation parameters $\gamma_\text{sim}^*$ and $\xi_\text{sim}^*$ but with the same value of $\xi^*$ ($\xi^*=1.26$ for disks and $\xi^*=1.68$ for spheres). The solid and dashed lines are the values obtained for \vicente{$a_3$} by means of Approximation I and Approximation II, respectively.
\label{fig3}}
\end{figure*}
\begin{figure*}
\begin{center}
\begin{tabular}{lr}
\resizebox{7.5cm}{!}{\includegraphics{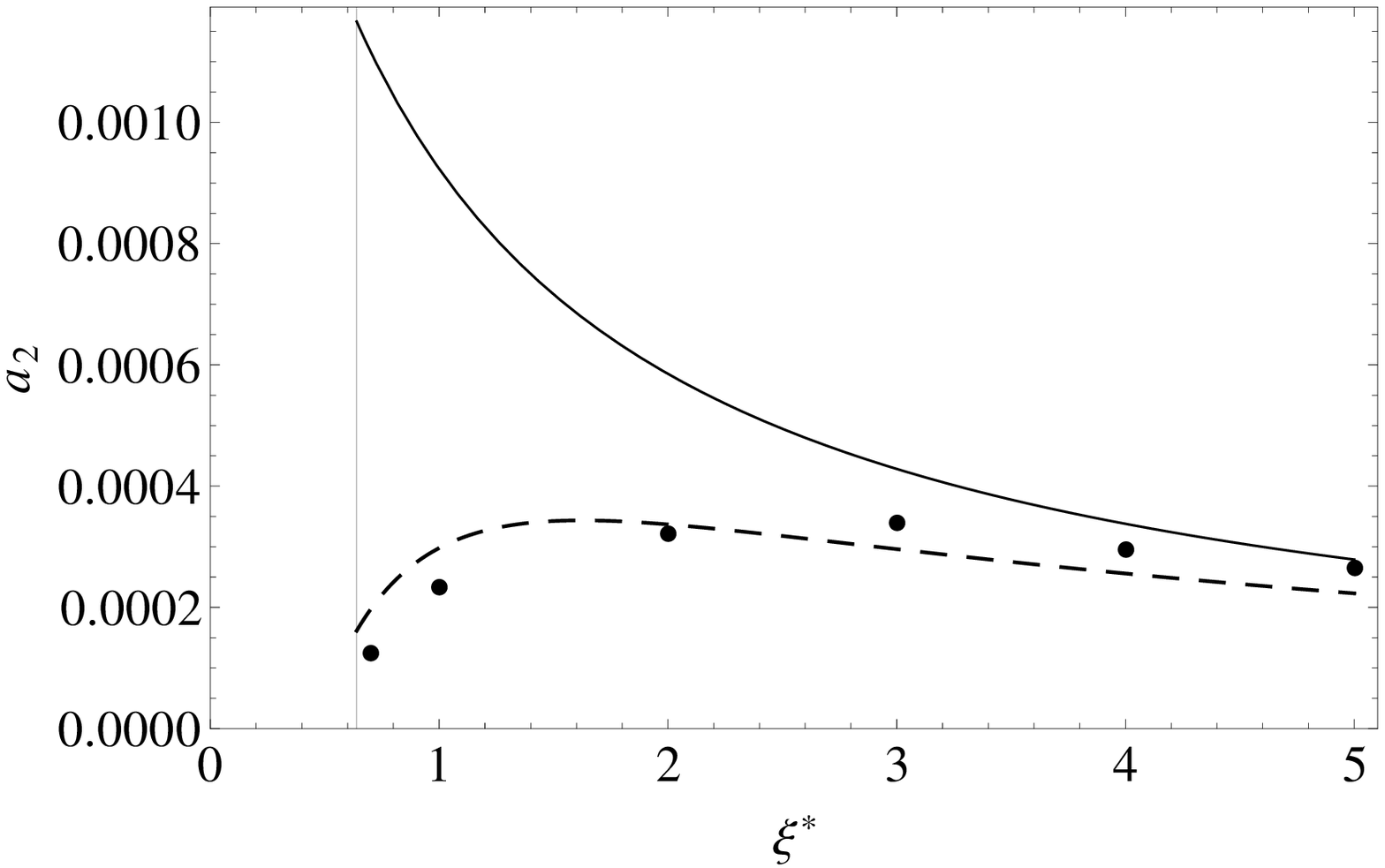}}&\resizebox{7.5cm}{!}
{\includegraphics{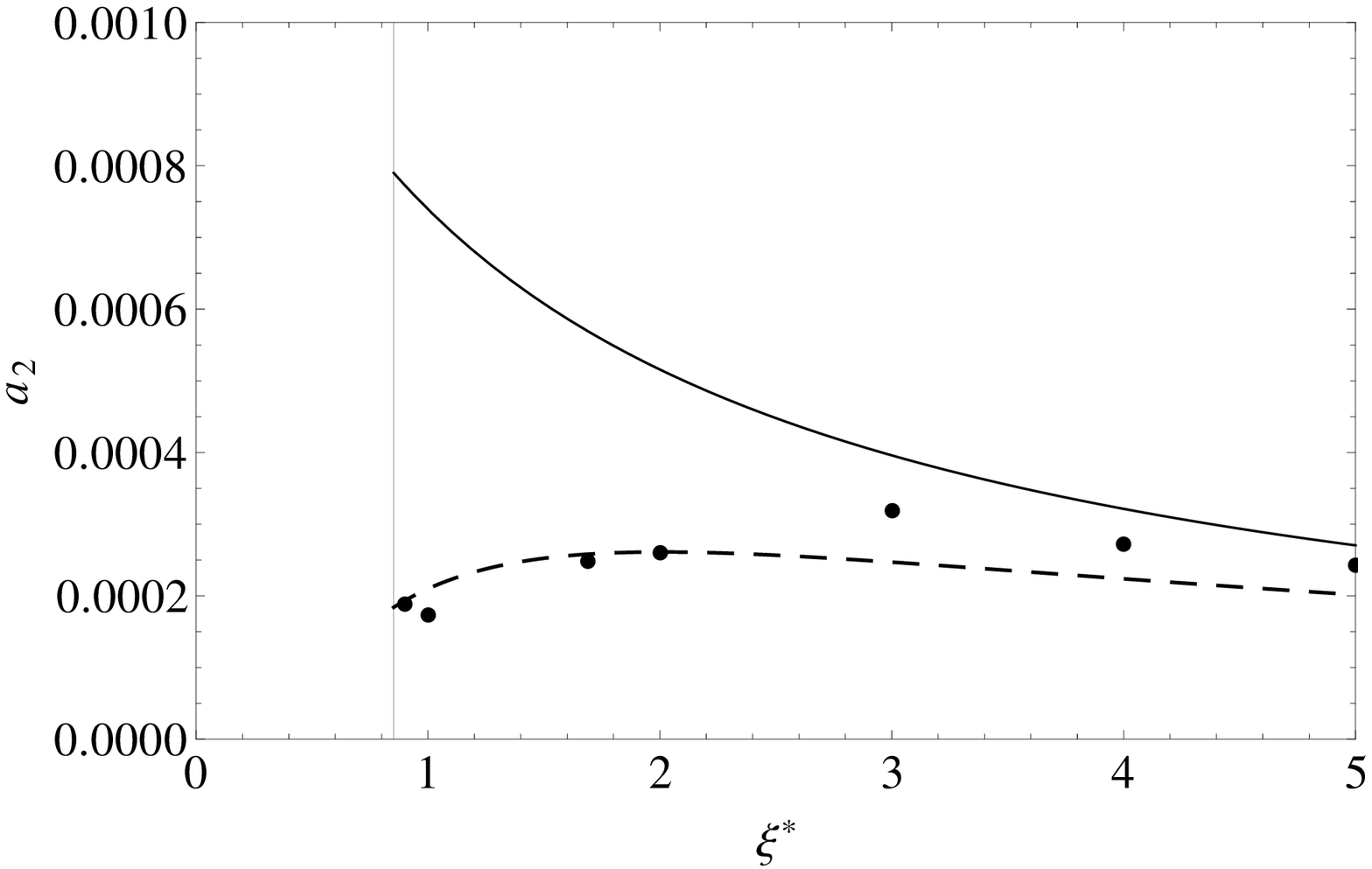}}
\end{tabular}
\end{center}
\caption{Plot of the second Sonine coefficient $a_2$ versus the (reduced) noise strength $\xi^*$ for $\al=0.7$ in the case of hard disks (left panel) and hard spheres (right panel). The symbols refer to simulation results while the solid and dashed lines are the values obtained for $a_2$ by means of Approximation I and Approximation II, respectively. The vertical lines indicate the threshold values $\xi^*_\text{th}$.
\label{fig4}}
\end{figure*}
\begin{figure*}
\begin{center}
\begin{tabular}{lr}
\resizebox{7.5cm}{!}{\includegraphics{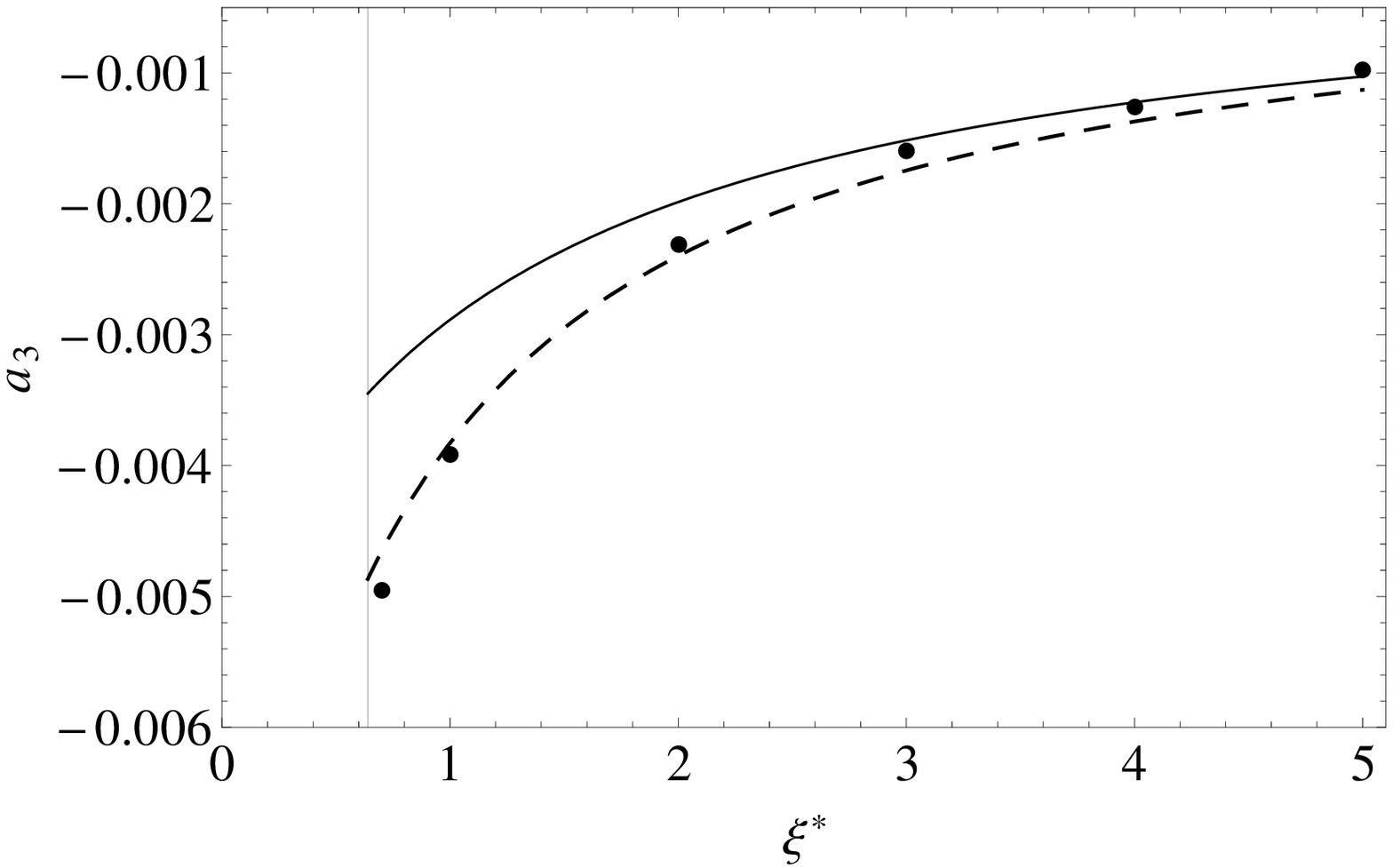}}&\resizebox{7.5cm}{!}
{\includegraphics{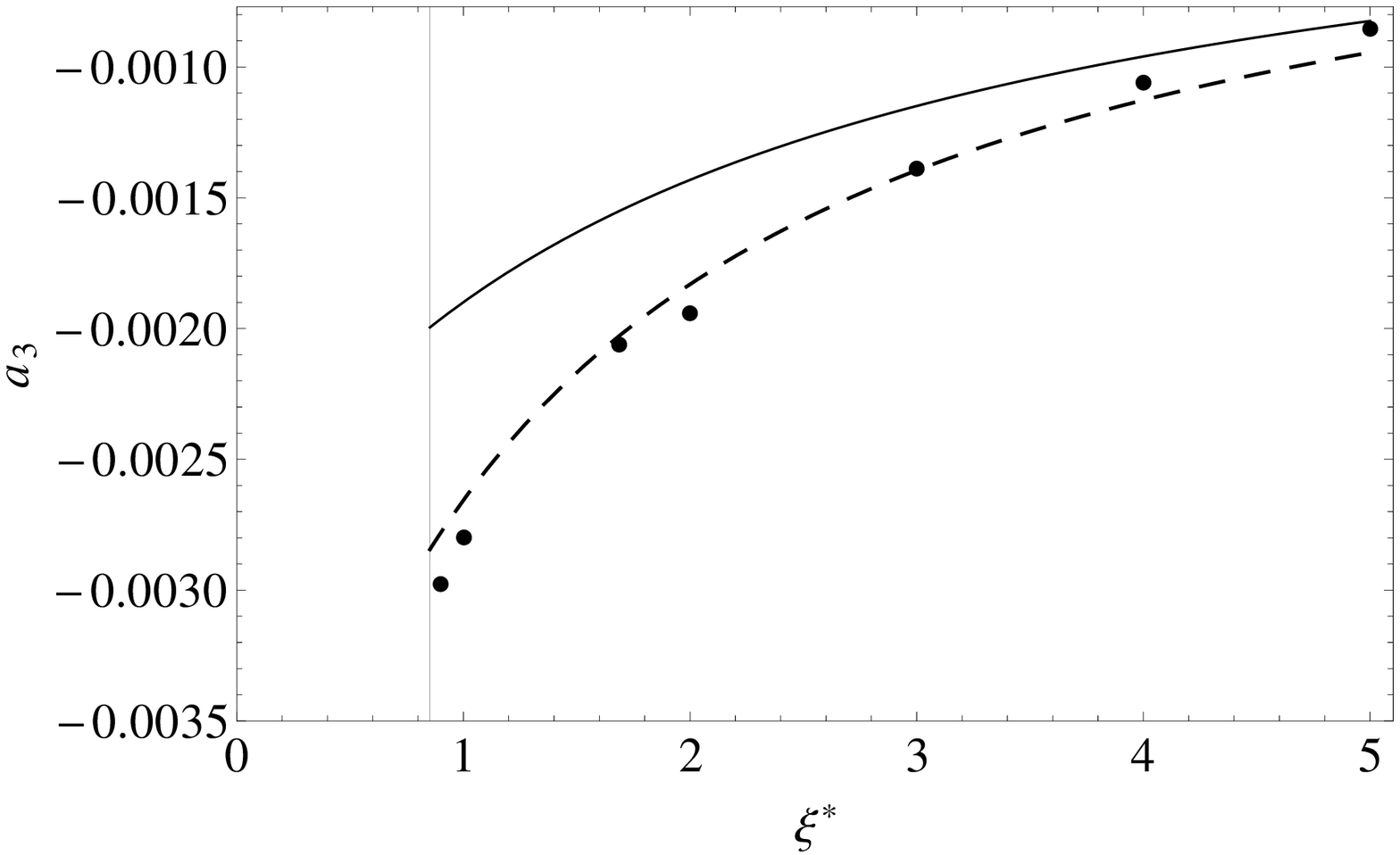}}
\end{tabular}
\end{center}
\caption{Plot of the third Sonine coefficient $a_3$ versus the (reduced) noise strength $\xi^*$ for $\al=0.7$ in the case of hard disks (left panel) and hard spheres (right panel). The symbols refer to simulation results while the solid and dashed lines are the values obtained for $a_3$ by means of Approximation I and Approximation II, respectively. The vertical lines indicate the threshold values $\xi^*_\text{th}$.
\label{fig5}}
\end{figure*}

The DSMC algorithm is composed in its basic form of a collision step, that takes care of all particle collisions, and a free drift step between particle collisions \cite{B94}. If volume forces act on the system, their corresponding steps need to be incorporated to the algorithm. \vicente{Although the DSMC method has been explained elsewhere \cite{B94, MS00,GCV13}, we will give here some details of the specific method we have used to solve the uniform Enskog-Boltzmann equation \eqref{2.7}.

The velocity distribution function is represented by the velocities $\left\{\mathbf{v}_i \right\}$ of $N$ ``simulated'' particles:
\beq
\label{f}
f(\mathbf{v},t)\to \frac{n}{N}\; \sum_{i=1}^N\; \delta (\mathbf{v}_i(t)-\mathbf{v}).
\eeq
The system is always initialized with a Maxwellian velocity distribution with temperature $T_0$. In the collision stage, a sample of $\frac{1}{2}N\omega_\text{max} dt$ pairs is chosen at random with equiprobability, where $dt$ is the time step (which is much smaller than the mean free time) and $\omega_\text{max}$ is an upper bound estimate of the probability that a particle collides per unit of time. For each pair $(i,j)$ belonging to this sample, a given direction $\widehat{\boldsymbol {\sigma}}_{ij}$ is chosen at random with equiprobability. Then, the collision between particles $i$ and $j$ is accepted with a probability equal to $\Theta (\mathbf{g}_{ij}\cdot \widehat{\boldsymbol {\sigma}}_{ij})\omega_{ij}/\omega_\text{max}$ where $\omega_{ij}=(4\pi n \sigma^2 \chi)|\mathbf{g}_{ij}\cdot \widehat{\boldsymbol {\sigma}}_{ij}|$ for hard spheres and $\omega_{ij}=(2\pi n \sigma \chi)|\mathbf{g}_{ij}\cdot \widehat{\boldsymbol {\sigma}}_{ij}|$ for hard disks. Here, $\mathbf{g}_{ij}=\mathbf{v}_i-\mathbf{v}_j$ is the relative velocity. If the collision is accepted, postcollisional velocities are assigned according to the scattering rule \eqref{2.3}. In the case that $\omega_{ij}>\omega_\text{max}$, then the estimate $\omega_\text{max}$ is updated as $\omega_\text{max}=\omega_{ij}$. Thus, notice that the acceptance probability $\Theta (\mathbf{g}_{ij}\cdot \widehat{\boldsymbol {\sigma}}_{ij})\omega_{ij}/\omega_\text{max}$ is independent of the pair correlation function and for this reason the DSMC algorithm is formally identical for both Boltzmann and Boltzmann-Enskog equations, if the system is homogeneous, like in our case \cite{MS00}.}

In the streaming stage, the velocity of every particle is changed according to the thermostat which is composed by two different forces. These two forces act consecutively (the precedence is not relevant) to the collision step. We only need to take care that the intrinsic time scales produced by the two forces ($\tau_\text{drag}=m/\gamma_\text{b}$ and $\tau_\text{st}=v_0^2/\xi_\text{b}$) are not too fast compared to the algorithm time step $dt$, which needs to be small compared to the characteristic collision time in order to describe properly the collision integral of the Boltzmann-Enskog equation \cite{B94,GCV13}. In other words, we need that $\tau_\text{drag}\le \nu^{-1}$ and $\tau_\text{st}\le \nu^{-1}$, where $\nu=v_0/\ell$.
As said in section \ref{sec2}, our thermostat is constituted by a deterministic external force proportional to the velocity particle plus a stochastic force. Consequently, the thermostat updates particle velocities following the rule
\begin{equation}
  \label{updateF}
  \mathbf{v}_i\to\mathbf{v}_i+\mathbf{w}_i^{\text{th}}, \quad \mathbf{w}_i^{\text{th}}=\mathbf{w}_i^{\text{drag}}+\mathbf{w}_i^{\text{st}}.
\end{equation}
Here, $\mathbf{w}_i^{\text{drag}}$ and $\mathbf{w}_i^{\text{st}}$ denote the velocity increments due to the drag and stochastic forces, respectively. The increment $\mathbf{w}_i^{\text{st}}$ is picked from a Gaussian distribution with a variance characterized by the noise intensity $\xi_\text{b}^2$ fulfilling the conditions
\begin{equation}
  \label{fstcond}
  \left<\mathbf{w}_i\right>=\mathbf{0}, \quad \left<\mathbf{w}_i\mathbf{w}_j\right>=\xi_b^2dt\delta_{ij},
\end{equation}
where
\begin{equation}
  \label{fst}
  P(w_i^{\text{st}})=(2\pi\xi_b^2dt)^{-3/2}e^{-{w_i^{\text{st}}}^2/(2\xi_b^2dt)}
\end{equation}
is a Gaussian probability distribution \cite{MS00}. The velocity increment $\mathbf{w}_i^{\text{drag}}$ due to the drag force is given by
\begin{equation}
  \label{fdrag}
  \mathbf{w}_i^{\text{drag}}=-\gamma_\text{b}\mathbf{v}_idt.
\end{equation}

In the simulations carried out in this work we have used a number of particles $N=2\times 10^6$ particles and a time step $dt=5\times 10^{-2}\nu_0^{-1}$, where $\nu_0^{-1}=(2T_0/m)^{1/2}n\sigma^{d-1}$. Moreover, for the sake of convenience, we introduce the following dimensionless quantities ($\gamma_\text{sim}^*$ and $\xi_\text{sim}^*$) characterizing the driven parameters used for the different simulations
\beq
\label{sim1}
\gamma_\text{sim}^*=\frac{\gamma_\text{b}}{\chi m \nu_0}=\left(\frac{T_\text{s}}{T_0}\right)^{1/2}\gamma^*,
\eeq
\beq
\label{sim2}
\xi_\text{sim}^*=\frac{m\xi_\text{b}^2}{\chi T_0 \nu_0}=\left(\frac{T_\text{s}}{T_0}\right)^{3/2}\xi^*.
\eeq
The last equality in equations \eqref{sim1} and \eqref{sim2} provides the relation between the simulation (reduced) quantities $\gamma_\text{sim}^*$ and $\xi_\text{sim}^*$ and their corresponding theoretical ones $\gamma^*$ and $\xi^*$, respectively.
\begin{figure*}
\begin{center}
\begin{tabular}{lr}
\resizebox{7.5cm}{!}{\includegraphics{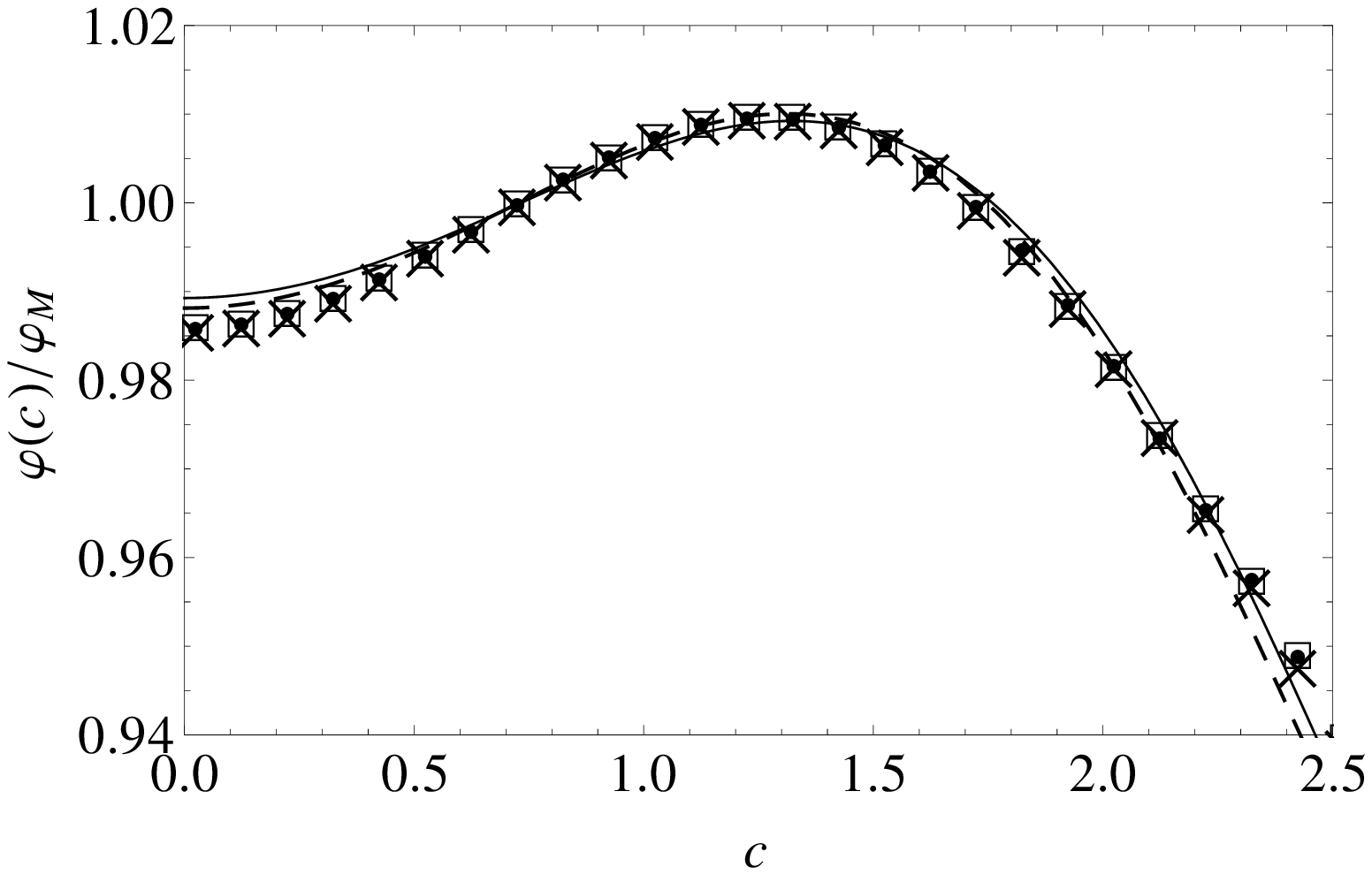}}&\resizebox{7.5cm}{!}
{\includegraphics{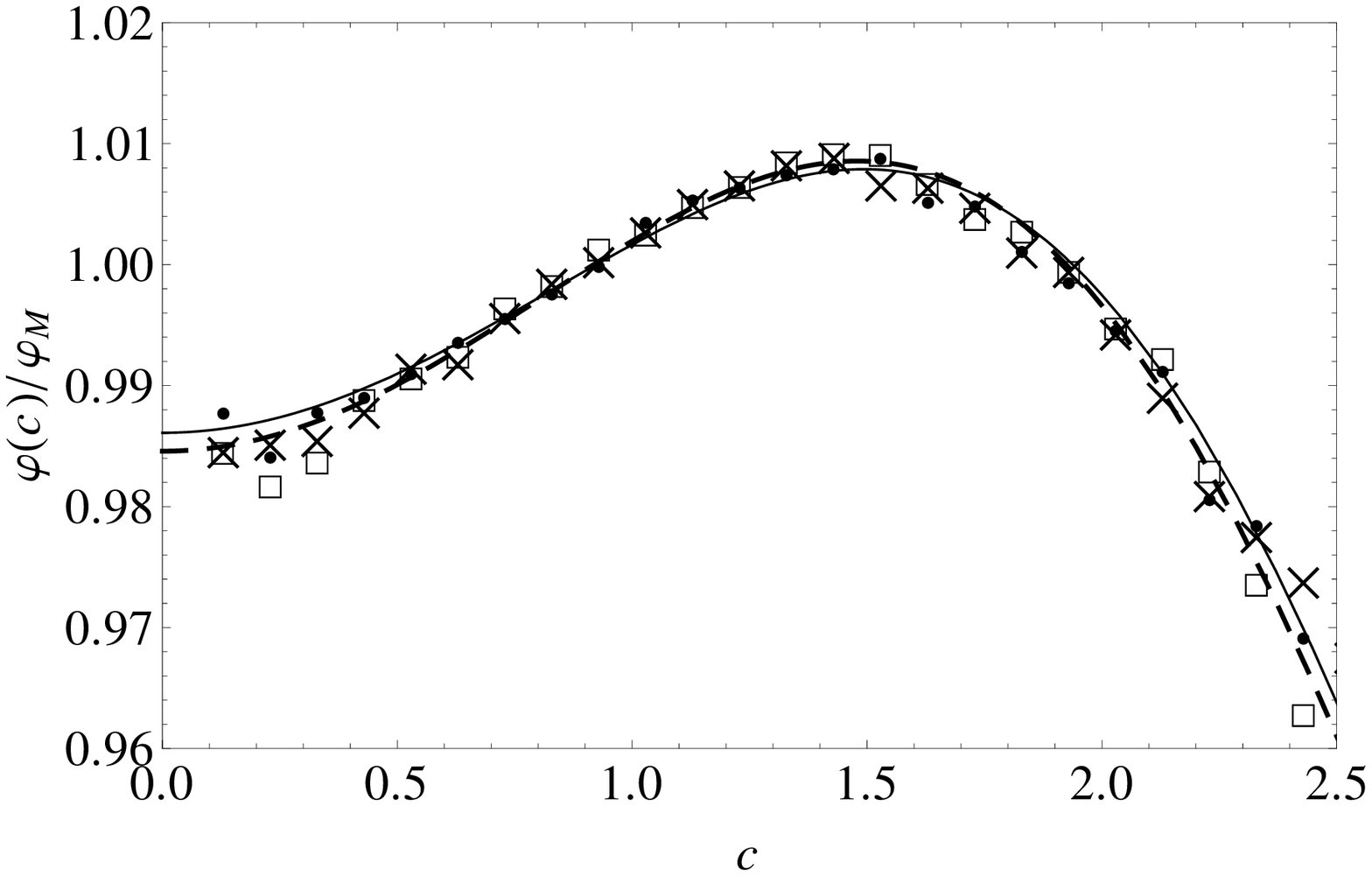}}
\end{tabular}
\end{center}
\caption{Plot of the scaled distribution function $\varphi(c,\xi^*)/\varphi_\text{M}(c)$ in the steady state for $\al=0.8$. The left panel is for hard disks while the right panel corresponds to hard spheres. The symbols refer to DSMC data obtained for three different systems with parameters: $\{\gamma_\text{sim}^*, \xi_\text{sim}^*\}=\{(1.4\times 10^{-2}, 5.2\times 10^{-5}), (9.8\times 10^{-3}, 1.8\times 10^{-5}), (7\times 10^{-3}, 6.5\times 10^{-6})\}$ for $d=2$ and $\{\gamma_\text{sim}^*, \xi_\text{sim}^*\}=\{(7.1\times 10^{-3}, 2.9\times 10^{-6}), (5\times 10^{-3}, 9.8\times 10^{-7}), (3.6\times 10^{-3}, 3.6\times 10^{-7})\}$ for $d=3$. These values yield a common value of $\xi^*$: $\xi^*=1.263$ for $d=2$ and $\xi^*=1.688$ for $d=3$. The lines correspond to equation \eqref{4.1} with expressions for the cumulants given by Approximation I (solid lines) and Approximation II (dashed lines).
\label{fig6}}
\end{figure*}
\begin{figure*}
\begin{center}
\begin{tabular}{lr}
\resizebox{7.5cm}{!}{\includegraphics{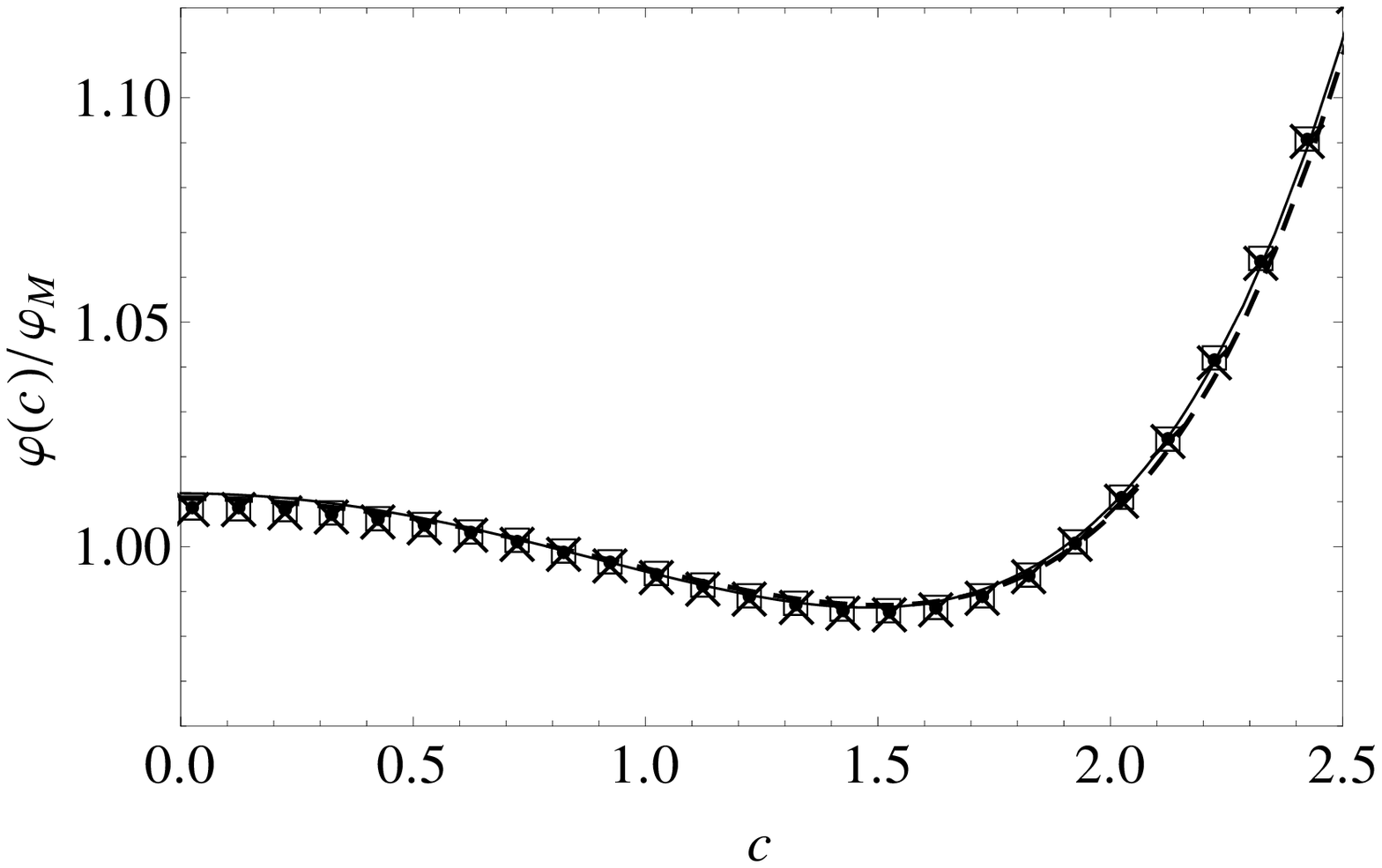}}&\resizebox{7.5cm}{!}
{\includegraphics{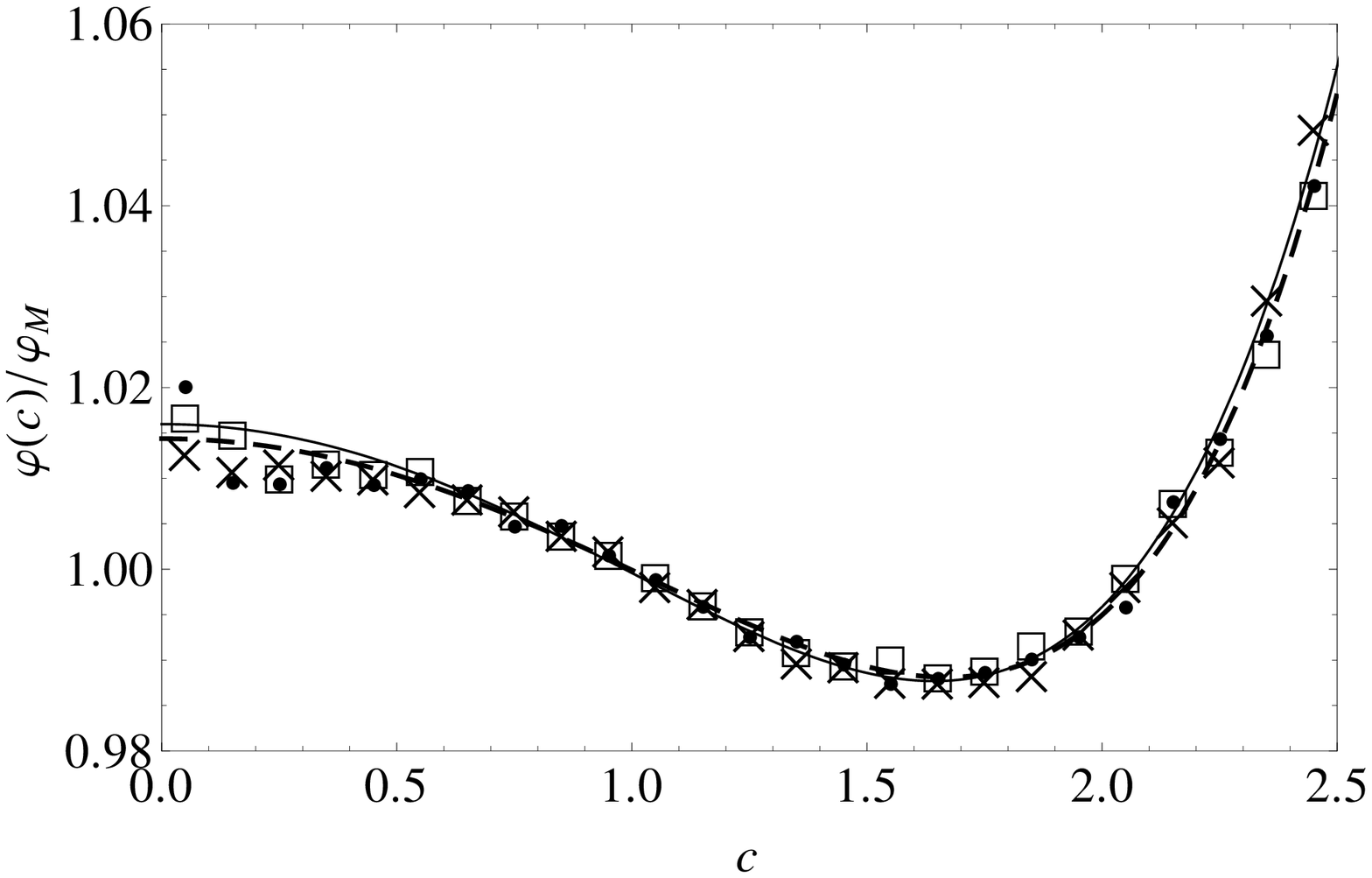}}
\end{tabular}
\end{center}
\caption{Plot of the scaled distribution function $\varphi(c,\xi^*)/\varphi_\text{M}(c)$ in the steady state for $\al=0.6$. The left panel is for hard disks while the right panel corresponds to hard spheres. The symbols refer to DSMC data obtained for three different systems with parameters: $\{\gamma_\text{sim}^*, \xi_\text{sim}^*\}=\{(1.4\times 10^{-2}, 2.9\times 10^{-4}), (9.8\times 10^{-3}, 10^{-4}), (7\times 10^{-3}, 3.6\times 10^{-5})\}$ for $d=2$ and $\{\gamma_\text{sim}^*, \xi_\text{sim}^*\}=\{(7.1\times 10^{-3}, 1.5\times 10^{-5}), (5\times 10^{-3}, 5.4\times 10^{-6}), (3.6\times 10^{-3}, 1.9\times 10^{-6})\}$ for $d=3$. These values yield a common value of $\xi^*$: $\xi^*=1.263$ for $d=2$ and $\xi^*=1.688$ for $d=3$. The lines correspond to equation \eqref{4.1} with expressions for the cumulants given by Approximation I (solid lines) and Approximation II (dashed lines).
\label{fig7}}
\end{figure*}

\subsection{Comparison between theory and simulations}

Although we are mainly interested in evaluating all the relevant quantities of the problem ($a_2$, $a_3$ and $\varphi$) in the (asymptotic) steady state, it is also interesting to analyze the approach of some of these quantities towards the steady state. Figure \ref{fig1} shows the time evolution of both the (reduced) temperature $T(t)/T_\text{s}$ (left panel) and the distribution function $\varphi(c_0)$ (right panel) for the (dimensionless) velocity $c_0=v_{0,\text{s}}/v_0(t)$. Here, $T_\text{s}$ and $v_{0,\text{s}}=\sqrt{2T_\text{s}/m}$ refer to the theoretical steady values of the granular temperature and thermal velocity, respectively. The solid horizontal lines correspond to the theoretical predictions by considering the first two non-Gaussian corrections (third Sonine approximation) to the distribution $\varphi$ (see equation \eqref{4.1}). We have made runs of identical systems except that they are initialized with different temperatures. After a transient regime, as expected we observe that all simulation data tend to collapse to the same steady values for sufficiently long times. In addition, the corresponding steady values obtained from the simulation
for both temperature and distribution function practically coincide with those predicted by the Sonine solution. It is also to be noticed that the convergence to the steady values occurs approximately at the same time for both $T(t)/T_\text{s}$ and $\varphi(c_0)$ (thermal fluctuations make difficult to determine the exact point for steady state convergence for the distribution function). This is another and indirect way of checking that indeed the normal solution exists for simulations, since its existence implies, from equation \eqref{fhd}, that we reach the scaled form \eqref{1.1} when the temperature is stationary.

\vicente{Some previous works on a granular gas heated by the stochastic thermostat \cite{GMT12} and on the simple shear flow \cite{AS07} have shown that before reaching the steady state the system evolves towards a universal \emph{unsteady} state that depends on a new parameter measuring the distance to the steady state. A similar behavior is expected here where the different solutions to the Enskog-Boltzmann equation \eqref{2.7} would be attracted by the universal distribution function $f(v,t)\to n v_0(t)^{-d} \varphi (\mathbf{c},\tilde{\gamma}(t),\tilde{\xi}(t))$, where $\mathbf{c}=\mathbf{v}/v_0(t)$, and
\beq
\label{4.0}
\tilde{\gamma}(t)\equiv \frac{\ell \gamma_\text{b}}{\chi m v_0(t)}, \quad
\tilde{\xi}(t)\equiv \frac{\ell \xi_\text{b}^2}{\chi T(t) v_0(t)}.
\eeq
The dimensionless driven parameters $\tilde{\gamma}(t)$ and $\tilde{\xi}(t)$ measure the distance to the steady state. Of course, for asymptotically long times, the steady state is eventually reached, i.e., $\varphi(\mathbf{c},\tilde{\gamma}(t),\tilde{\xi}(t))\to \varphi_\text{s}(\mathbf{c},\xi^*)$, where $\xi^*$ is defined by Eq.\ \eqref{2.13}. The above unsteady hydrodynamic regime (for which the system has forgotten its initial condition) is expected to be achieved after a certain number of collisions per particle. On the other hand, although the characterization of this unsteady state is a very interesting problem, its study lies beyond the goal of the present paper.}

Now, we will focus on the steady state values of the relevant quantities of the problem. In particular,
the basic quantities measuring the deviation of the distribution function from its Maxwellian form are the second and third Sonine coefficients $a_2$ and $a_3$, respectively. The dependence of $a_2$ and $a_3$ on the coefficient of restitution $\al$ is shown in figures \ref{fig2} and \ref{fig3}, respectively, for hard disks (left panels) and spheres (right panels). Three different systems with different values of the simulation parameters $\gamma_\text{sim}^*$ and $\xi_\text{sim}^*$ but with the same value of $\xi^*$ ($\xi^*=1.263$ for disks and $\xi^*=1.688$ for spheres) have been considered. We observe that, at a given of $\al$, the corresponding three simulation data collapse in a common curve, showing that indeed both Sonine coefficients are always of the form $a_i(\alpha, \xi^*)$. Regarding the comparison between theory and simulation, it is quite apparent that while both Approximations I and II compare quantitatively quite well with simulations in the case of $a_2$, Approximation II has a better performance than Approximation I in the case of $a_3$, specially at very strong dissipation. This is the expected result since Approximation II is in principle more accurate that Approximation I, although the latter is more simple than the former. In this sense and with respect to the $\al$-dependence of $a_2$ and $a_3$, Approximation I could be perhaps preferable to Approximation II since it has an optimal compromise between simplicity and accuracy.

On the other hand, more quantitative discrepancies between both Approximations are found when one analyzes both Sonine coefficients vs. $\xi^*$ with constant $\al$. Figures \ref{fig4} and \ref{fig5} show $a_2$ and $a_3$, respectively, versus $\xi^*$ at $\al=0.7$. We see that Approximation I exhibits a poor agreement with simulations since it predicts a dependence on the noise strength opposite to the one found in the simulations. On the other hand, Approximation II agrees very well with simulation data in all the range of values of $\xi^*$  (note that $\xi^*\gtrsim 0.639$ for $d=2$ and $\xi^*\gtrsim 0.852$ for $d=3$ to achieve a steady state for $\al=0.7$). It must be also noted that for the systems studied in figures \ref{fig4} and \ref{fig5}, although the magnitudes of both Sonine coefficients are very small, $|a_2|$ is of the order of ten times smaller than $|a_3|$. \vicente{This may indicate that in certain ranges the cumulant $a_3$ is relevant compared to $a_2$, which justifies our Approximation II.}

The small values of the coefficients $a_2$ and $a_3$ support the assumption of a low-order truncation in polynomial expansion and suggests that the scaled distribution function $\varphi(c,\xi^*)$ for thermal velocities can be well represented by the three first contributions (note that $a_1=0$) in the Sonine polynomial expansion \eqref{3.1}. To confirm it, we have measured the deviation of $\varphi(c,\xi^*)$ from its Maxwellian form $\varphi_\text{M}(c)$. In figures \ref{fig6} an \ref{fig7} we plot the ratio $\varphi(c,\xi^*)/\varphi_\text{M}(c)$ versus the reduced velocity $c$ in the steady state for two values of the coefficient of restitution ($\al=0.8$ and $\al=0.6$). As before, we have considered a system of inelastic hard disks (figure \ref{fig6} with $\xi^*=1.26$) and inelastic hard spheres (figure \ref{fig7} with $\xi^*=1.69$). As in figures \ref{fig2}--\ref{fig5}, symbols correspond to simulation results obtained for different values of $\gamma_\text{sim}^*$ and $\xi_\text{sim}^*$. The solid and dashed lines are obtained from equation \eqref{3.1} with the series truncated at $p=3$, i.e.,
\beqa
\label{4.1}
\frac{\varphi(c,\xi^*)}{\varphi_\text{M}(c)}&\to& 1+a_2(\xi^*)\left(\frac{1}{2}c^4-\frac{d+2}{2}c^2+\frac{d(d+2)}{8}\right)\nonumber\\
& & -a_3(\xi^*)\left(
\frac{1}{6}c^6-\frac{d+4}{4}c^4+\frac{(d+2)(d+4)}{8}c^2-\frac{d(d+2)(d+4)}{48}\right).
\eeqa
The coefficients $a_2$ and $a_3$ in equation \eqref{4.1} are determined by using Approximation I (solid lines) and Approximation II( dashed lines). First, it is quite apparent that simulations confirm that the reduced distribution function $\varphi(c,\xi^*)$ is a universal function of $\xi^*$ since all simulation series at constant $\xi^*$ collapse to the same curve (within non-measurable margin error). We also see that the simulation curves agree very well with the corresponding third-degree Sonine polynomial in this range of velocities, especially in the two-dimensional case. Surprisingly, in the high velocity region, the curves obtained from Approximation I fits the simulation data slightly better than those obtained by using the improved Approximation II. In any case, the agreement between theory and simulation is again excellent, especially taking into account the very small discrepancies we are measuring.


\section{Concluding remarks}
\label{discussion}

In this paper we have performed Monte Carlo simulations of the Enskog-Boltzmann for a granular fluid in a homogeneous state. The system is driven by a stochastic bath with friction. One of the primary objectives of this work has been to check the velocity scaling and form assumed for the distribution function  in the steady state. As equation \eqref{1.1} indicates, the new feature of the scaled distribution $\varphi$ is that it not only depends on the granular temperature $T$ through the (scaled) velocity $c$ but also through the (reduced) noise strength $\xi^*$ (defined in equation \eqref{2.13}). The simulation results reported here (see figures \ref{fig6} and \ref{fig7}) have confirmed the above dependence since different systems sharing the same values of $\xi^*$ and $\al$ lead to the same distribution function $\varphi$. This is consistent with the existence of a \emph{normal} solution in the long-time limit.

Apart from performing Monte Carlo simulations to confirm the validity of a hydrodynamic description for finite degree of collisional dissipation, we have also characterized the distribution $\varphi$ through its first velocity moments. More specifically, we have obtained the second $a_2$ and third $a_3$ Sonine coefficients. While the coefficient $a_2$ measures the fourth-degree velocity moment of $\varphi$, the coefficient $a_3$ is defined in terms of the sixth-degree velocity moment of $\varphi$. Both Sonine coefficients provide information on the deviation of $\varphi$ from its Maxwellian form $\varphi_\text{M}$. Moreover, the knowledge of those coefficients are important, for instance, in the precise determination of the transport coefficients \cite{GCV13}. On the other hand, given that the Sonine coefficients cannot be \emph{exactly} determined (they obey an infinite hierarchy of moments), one has to truncate the corresponding Sonine polynomial expansion in order to estimate them. Here, we have considered two different approaches (Approximation I and II) to get explicit expressions of $a_2$ and $a_3$ in terms of the dimensionality of the system $d$, the coefficient of restitution $\al$ and the driven parameter $\xi^*$. Approximation II is more involved than Approximation I since it considers both Sonine coefficients as being of the same order of magnitude.  The comparison between the analytical solution and DSMC results shows in general a good agreement, even for high-inelasticity. Moreover, as expected, the improved Approximation II for $a_2$ and $a_3$ shows a better agreement with simulations than Approximation I (see figures \ref{fig2}--\ref{fig5}). Thus, taking into account all the above comparisons, we can conclude that a good compromise between accuracy and simplicity is represented by  Approximation I.

The results derived in this paper show clearly that the combination of analytical and computational tools (based on the DSMC method) turns out to be an useful way to characterize properties in granular flows. On the other hand, given that most of the Sonine coefficients could be directly measured by DSMC, one could in principle make a least-square fit to obtain explicit forms for those coefficients. However, this procedure would not be satisfactory from a more fundamental point of view, especially if one is interested in capturing the behavior of $\varphi(c)$ and its Sonine expansion. In this context, our analytical solution of the distribution function (redundant as it may seem) has the advantage of providing a rational description of the physical properties of the kinetic equation of the system. This is not accomplished by the numerical solution. However, the fact that the DSMC method gives an accurate numerical solution of the Enskog-Boltzmann equation makes it complementary to the theoretical one and thus both conform a complete description of the kinetic equation of our system.

\acknowledgments

The present work has been supported by the Ministerio de
Educaci\'on y Ciencia (Spain) through grants No. FIS2010-16587 (V.G., M.G.Ch. and F.V.) and No. MAT2009-14351-C02-02 (F.V.). The first Grant has been partially financed by
FEDER funds and by the Junta de Extremadura (Spain) through Grant No. GRU10158. The  research  of M. G. Chamorro  has  been  supported  by  the  predoctoral  fellowship BES-2011-045869 from the  Spanish Government (Spain).

\appendix
\section{Expressions for $A_i$, $B_i$, and $C_i$}
\label{appA}

In this Appendix we provide the explicit expressions of the coefficients $A_i$, $B_i$, and $C_i$ as functions of $d$ and $\alpha$. They are given by \cite{NE98,BP06}
\begin{equation}
A_0=K(1-\alpha^2),\quad A_2=\frac{3K}{16}(1-\alpha^2),\quad
A_3=\frac{K}{64}(1-\alpha^2),
\label{a1}
\end{equation}
\begin{equation}
B_0=K(1 -\alpha^2) \left(d + \frac{3}{2} + \alpha^2\right),
\end{equation}
\begin{equation}
B_2=K(1+\alpha)\left[d -1+\frac{3}{32} (1 - \alpha) (10 d + 39 + 10 \alpha^2)
\right],
\end{equation}
\begin{equation}
B_3=-\frac{K}{128}(1+\alpha)\left[(1 - \alpha) (97 + 10 \alpha^2) +2(d -
1)(21 - 5 \alpha)\right],
\end{equation}
\begin{equation}
C_0=
 \frac{3K}{4} (1 - \alpha^2) \left[d^2 +
    \frac{19}{4}+(d + \alpha^2) (5 + 2 \alpha^2)\right],
\end{equation}
\begin{equation}
C_2=
 \frac{3K}{256} (1 - \alpha^2) \left[1289 + 172 d^2+
     4 (d + \alpha^2) (311 + 70 \alpha^2) \right] +
  \frac{3}{4} {\lambda},
\end{equation}
\begin{equation}
C_3= -\frac{3K}{1024} (1 - \alpha^2) \left[2537 + 236 d^2+
     4 (d + \alpha^2) (583 + 70 \alpha^2)\right] -
  \frac{9}{16} \lambda,
\end{equation}
where
\begin{equation}
K\equiv \frac{\pi^{(d-1)/2}}{\sqrt{2}\Gamma(d/2)},\quad
\lambda\equiv K(1 + \alpha) \left[(d - \alpha) (3 +
       4 \alpha^2) + 2 (d^2 - \alpha)\right].
\end{equation}

\section{Approximations I and II}
\label{appB}

In this Appendix we display the forms of the Sonine coefficients $a_2$ and $a_3$ by using Approximations I and II. Let us start by considering Approximation I. In this case, we neglect $a_3$ versus $a_2$ in equation\ \eqref{3.18} and so, one gets
\begin{equation}
\label{b1}
\left[B_2-(d+2)(A_0+A_2)+\frac{d(d+2)}{2}\xi^*\right]a_2=(d+2)A_0-B_0,
\end{equation}
whose solution is
\begin{eqnarray}
\label{b2}
a_2^{(I)}(\alpha,\xi^*)&=&\frac{(d+2)A_0-B_0}{B_2-(d+2)(A_0+A_2)+\frac{d(d+2)}{2}\xi^*}\nonumber\\
&=&\frac{16(1-\alpha)(1-2\alpha^2)}{9+24d-\alpha(41-8d)+30(1-\alpha)\alpha^2
+\Omega_\text{d}\frac{\xi^*}{(1+\alpha)}},
\end{eqnarray}
where in the last step use has been made of the explicit expressions of $A_0$, $A_2$, $B_0$, and $B_2$. Here,
\beq
\label{b2.1}
\Omega_\text{d}=16\sqrt{2}d(d+2)\frac{\Gamma(d/2)}{\pi^{(d-1)/2}}.
\eeq
Once $a_2$ is determined, we can use equation\ \eqref{3.19} to express $a_3$ in terms of $a_2$. The result can be written as
\begin{equation}
\label{b3}
a_3^{(I)}(\alpha,\xi^*)=F\left(\alpha,a_2^{(I)}(\alpha),\xi^*\right),
\end{equation}
where
\begin{equation}
\label{b4}
F\left(\alpha,a_2,\xi^*\right)\equiv\frac{\frac{3}{4}(d+2)(d+4)A_0-C_0-\left[C_2+\frac{3}{4}(d+2)(d+4)
(d\xi^*-3A_0-A_2)\right]a_2}
{C_3-\frac{3}{4}(d+2)(d+4)
\left(A_3-A_0+\frac{d}{2}\xi^*\right)}.
\end{equation}


In Approximation II, $a_3$ is formally treated as being of the same order of magnitude as $a_2$ and so, Eqs.\ \eqref{3.18} and \eqref{3.19} become a linear set of two coupled equations for $a_2$ and $a_3$. The problem is algebraically more involved as in  Approximation I. The form of $a_2^{(II)}$ is given by equation \eqref{b5} where \begin{equation}
\label{b6}
M(\alpha,\xi^*)\equiv\left[C_3-\frac{3}{4}(d+2)(d+4)\left(A_3-A_0+\frac{d}{2}\xi^*\right)\right]
\left[(d+2)A_0-B_0\right]-\left[B_3-(d+2)A_3\right]\left[\frac{3}{4}(d+2)(d+4)A_0-C_0\right],
\end{equation}
and
\begin{eqnarray}
\label{b7}
N(\alpha,\xi^*)&\equiv& \left[B_2-(d+2)(A_0+A_2)+\frac{d(d+2)}{2}\xi^*\right]
\left[C_3-\frac{3}{4}(d+2)(d+4)\left(A_3-A_0+\frac{d}{2}\xi^*\right)\right]\nonumber\\
& &
-\left[B_3-(d+2)A_3\right]\left[C_2+\frac{3}{4}(d+2)(d+4)(d\xi^*-3A_0-A_2)\right].
\end{eqnarray}
The corresponding result for $a_3^{(II)}$ in approximation II has the same form as for approximation I except that now relies on $a_2^{(II)}$, i.e,
\begin{equation}
\label{b8}
a_3^{(II)}(\alpha,\xi^*)=F\left(\alpha,a_2^{(II)}(\alpha),\xi^*\right).
\end{equation}

\bibliography{HSD}

\end{document}